%% file: main.tex
\documentclass[sigconf]{acmart}

\AtBeginDocument{%
  }

\copyrightyear{2026}
\acmYear{2026}
\setcopyright{cc}
\setcctype{by}
\acmConference[VRST '26]{32nd ACM Symposium on Virtual Reality Software and Technology}{November 16--18, 2026}{Sendai, Japan}
\acmBooktitle{32nd ACM Symposium on Virtual Reality Software and Technology (VRST '26), November 16--18, 2026, Sendai, Japan}
\acmDOI{10.1145/3822517.3848728}
\acmISBN{979-8-4007-2811-2/2026/11}

\usepackage{longtable}
\usepackage{array}

\makeatletter
\patchcmd{\@mkbibcitation}{\ref{TotPages}}{\pageref*{MainPaperLastPage}}{}{\PackageError{arxiv-combined}{Could not preserve main-paper page count}{Use the supplied acmart 2.16-compatible source.}}
\patchcmd{\@mkbibcitation}{\ref{TotPages}}{\pageref*{MainPaperLastPage}}{}{\PackageError{arxiv-combined}{Could not preserve main-paper page count}{Use the supplied acmart 2.16-compatible source.}}
\makeatother

\begin{document}

\title[Dream-Butterfly]{Dream-Butterfly: Configuring Embodied Conversational Guidance for Public Outdoor Mixed Reality Exhibitions}

% =========================
% Authors
% =========================

\author{Dongyijie Primo Pan}
\orcid{0009-0005-9830-1614}
\email{dpan750@connect.hkust-gz.edu.cn}
\affiliation{
  \department{Computational Media and Arts}
  \institution{The Hong Kong University of Science and Technology (Guangzhou)}
  \city{Guangzhou}
  \state{Guangdong}
  \country{China}
}

\author{Shuyue Li}
\orcid{0009-0003-5516-9372}
\email{sli688@connect.hkust-gz.edu.cn}
\authornote{These authors contributed equally to this work.}
\affiliation{
  \department{Computational Media and Arts}
  \institution{The Hong Kong University of Science and Technology (Guangzhou)}
  \city{Guangzhou}
  \state{Guangdong}
  \country{China}
}

\author{Yawei Zhao}
\orcid{0000-0001-9298-0955}
\email{yzhao099@connect.hkust-gz.edu.cn}
\authornotemark[1]
\affiliation{
  \department{Computational Media and Arts}
  \institution{The Hong Kong University of Science and Technology (Guangzhou)}
  \city{Guangzhou}
  \state{Guangdong}
  \country{China}
}

\author{Junkun Long}
\orcid{0009-0004-6705-6460}
\email{jlong892@connect.hkust-gz.edu.cn}
\authornotemark[1]
\affiliation{
  \department{Computational Media and Arts}
  \institution{The Hong Kong University of Science and Technology (Guangzhou)}
  \city{Guangzhou}
  \state{Guangdong}
  \country{China}
}

\author{Hao Li}
\orcid{0009-0007-2434-4325}
\email{hli307@connect.hkust-gz.edu.cn}
\authornotemark[1]
\affiliation{
  \department{Computational Media and Arts}
  \institution{The Hong Kong University of Science and Technology (Guangzhou)}
  \city{Guangzhou}
  \state{Guangdong}
  \country{China}
}

\author{Pan Hui}
\orcid{0000-0001-6026-1083}
\email{panhui@ust.hk}
\authornote{Corresponding author.}
\affiliation{
  \department{Computational Media and Arts}
  \institution{The Hong Kong University of Science and Technology (Guangzhou)}
  \city{Guangzhou}
  \state{Guangdong}
  \country{China}
}
\affiliation{
  \department{Division of Emerging Interdisciplinary Areas}
  \institution{The Hong Kong University of Science and Technology}
  \city{Clear Water Bay, Kowloon}
  \country{Hong Kong}
}

\renewcommand{\shortauthors}{Pan et al.}

% =========================
% Abstract
% =========================

\begin{abstract}

Public outdoor mixed reality (MR) exhibitions make guidance a spatial interaction problem: visitors need timely explanations while roaming open sites, attending to virtual artworks, hazards, bystanders, and route choices. We present Dream-Butterfly, a field-deployed embodied conversational guide for campus-scale outdoor MR exhibitions. Visitors explicitly summon the lightweight non-humanoid guide; it returns visibly to a hand-relative dialogue position and answers using retrieval-augmented responses scoped by the MR runtime to the artwork currently encountered. Within the resulting mixed guidance ecology, staff remain accountable for safety, wayfinding, device support, and contingencies. Dream-Butterfly was deployed with over 30 spatially anchored artworks across a 26,000~m$^2$ campus site, which we use as a stress case for public MR guidance under walking, field-of-view limits, route choice, bystander exposure, and safety constraints. In an in-the-wild study with 24 visitors, we compared two role arrangements: agent-default, where the embodied guide served as the default channel for artwork interpretation, and docent-default, where staff provided primary narration while the agent remained optional. Agent-default guidance was associated with more available, visitor-timed explanations, higher immersion/engagement, and stronger hedonic quality, while increasing visitors' self-curation burden around pacing, attention management, and question formulation. We argue that making interpretation available at visitors' chosen moments improves the timing of explanation in outdoor MR and shifts more work to visitors: deciding when to stop, what to ask, how deeply to engage, and when to move on.

\end{abstract}

% =========================
% CCS Concepts
% =========================



% =========================
% Keywords
% =========================


\maketitle

\begin{figure*}[t]
  \centering
  \includegraphics[width=\textwidth]{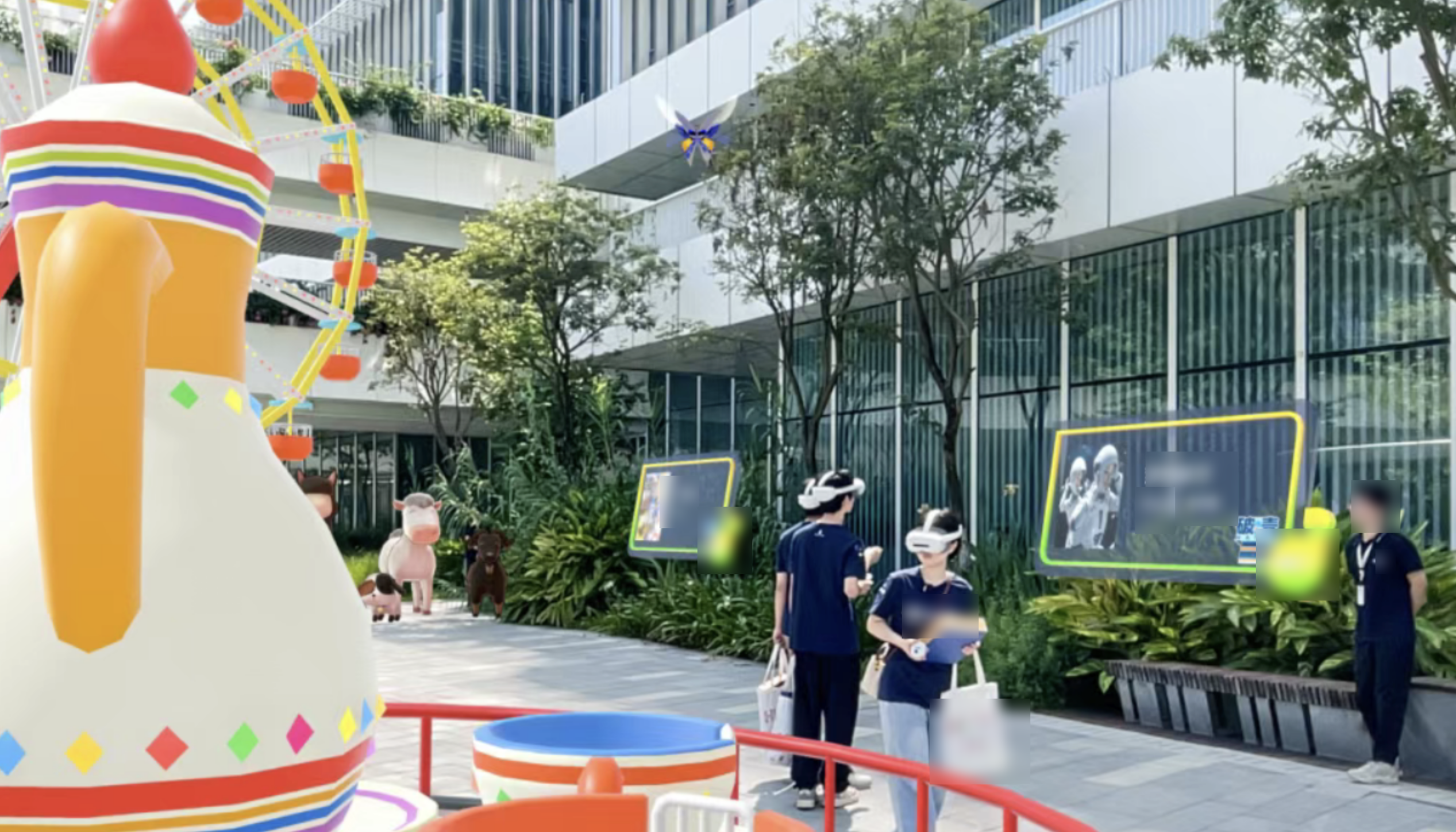}
  \caption{Campus-scale field deployment as a stress case for public outdoor MR guidance. Visitors explore a 26,000~m$^2$ open campus wearing HMD-based MR headsets while Dream-Butterfly remains nearby in an idle, non-intrusive state until explicitly summoned.}
  \Description{Outdoor campus courtyard during an MR art exhibition. Two visitors wearing head-mounted MR devices stand on a walkway near planted greenery and large display boards. A small butterfly-like virtual guide appears above them, hovering in the scene.}
  \label{fig:field-overview}
\end{figure*}

% =========================
% Paper Body
% =========================

\section{Introduction}

Public outdoor mixed reality (MR) exhibitions spread spatial interfaces across real sites. Visitors wearing head-mounted displays encounter virtual artworks while coordinating visual attention, walking, route choice, social exposure, and safety. Free roaming can support serendipitous public-space art experiences~\cite{knabe2025lucky,morse2021casual,10972673,harrison2025making,10.1145/3769534.3769544}, while dispersed artworks are encountered in different orders and at different speeds. The challenge is to make interpretation available when needed while preserving attention to public surroundings and physical hazards.

Existing guidance channels address different parts of this problem. Human docents provide adaptive explanation~\cite{best2012making} while carrying responsibility for wayfinding, troubleshooting, localisation failures, hazards, and other contingencies. Virtual labels registered to individual artworks, pre-authored audio, and route signage provide useful first-pass information; dynamic, multilingual, conceptually diverse, and freely ordered encounters also create demand for follow-up questions~\cite{roberts2018digital,raptis2021mumia}. Prior outdoor AR/MR touring systems show that information placement, timing, and environmental competition shape movement through place~\cite{matviienko2022arsightseeing,ghaemi2023placement}. What remains open is how artwork-specific explanation can be available at moments of curiosity while preserving the responsibilities that belong to human staff.

Conversational artificial intelligence (AI) offers one way to make situated explanation more flexible in extended reality (XR)~\cite{zhu2026genaixr,yu2026realitycopilot,sun2026jarvis,ihara2026cinemaworld}. A conversational guide's fit with a public outdoor visit depends on answer generation, conversation timing, spatial location, artwork-specific grounding, and the responsibilities retained by staff. As visitors turn, walk, and move between artworks, the guide can leave their field of view and the referent of a question can become unclear. In shared public space, casual speech and deliberate input can be difficult to distinguish, while visitors may still need staff for safety or device problems.

We address this challenge with \emph{Dream-Butterfly}, an embodied conversational guide designed for a campus-scale outdoor MR exhibition. Visitors select the encountered artwork with a controller ray, then summon the lightweight non-humanoid guide through a press-and-hold grip gesture. The butterfly returns and lands near the hand as a visible readiness cue; dialogue proceeds through speech and optional subtitles before visitors resume roaming. The guide follows in world coordinates and its retrieval-grounded responses are scoped by the MR runtime to the exhibition artwork currently encountered. Dream-Butterfly operates within a mixed guidance ecology where staff remain accountable for safety, wayfinding, device support, and contingencies.

We deployed Dream-Butterfly in a public outdoor MR exhibition on an open university campus. The exhibition featured over \textbf{30} spatially anchored MR artworks from creators across \textbf{9} countries and cultural backgrounds, distributed across roughly \textbf{26{,}000~m$^2$} of walkable campus space. This deployment provides a stress case for embodied conversational guidance under walking, changing field of view, public visibility, route choice, physical hazards, multilingual visitors, and mixed human--system support. In a between-subject field study with \textbf{24} visitors, we compared an \emph{agent-default} condition, where Dream-Butterfly was the first resort for artwork interpretation, with a \emph{docent-default} condition, where staff provided primary narration while Dream-Butterfly remained optional. The comparison examines how assigning a default interpretation channel shapes a mixed ecology of agent, staff, labels, and visitors.

Here, \emph{configuring} denotes the assignment of default responsibilities across the embodied agent, docents, labels, and visitors within this mixed guidance ecology.

This paper addresses two research questions:

\textbf{RQ1.} \emph{How does making a summonable embodied conversational agent the default source of artwork interpretation shape visitors' access to explanations, immersion/engagement, workload, and responsibility for pacing and question formulation?}

\textbf{RQ2.} \emph{How do visitors distribute guidance responsibilities among staff, the embodied agent, virtual labels, and themselves across interpretation, pacing, navigation, troubleshooting, and safety?}

\noindent\textbf{Contributions.}
This paper contributes to spatial user interface and XR guidance research in three ways. First, we contribute \emph{Dream-Butterfly} as a field-deployed interaction design for public outdoor MR guidance. Visitors invoke the lightweight non-humanoid guide with a press-and-hold gesture and see it return to a hand-relative dialogue position. Retrieval is limited to the selected artwork, and explicit cues show which requests remain with human staff. Together, these choices make interpretation available and re-acquirable during roaming, while keeping safety, wayfinding, device support, and contingencies with human staff. Second, we report a campus-scale field deployment with over 30 spatially anchored artworks across approximately 26,000~m$^2$ of open public campus space, contributing ecological evidence under walking, changing field of view, public visibility, route choice, physical hazards, multilingual visitors, and mixed human--system support. Third, through an in-the-wild comparison of two default-role conditions with 24 visitors, we show that agent-default guidance was associated with more visitor-timed access to explanation, higher immersion/engagement, and stronger hedonic quality, while also shifting more pacing, attention-management, and question-formulation work onto visitors.

\section{Related Work}

Interpretation in cultural venues is often understood as dialogic meaning-making that emphasises exchange over one-way information delivery~\cite{hein1999meaning,reynolds2011forum,cameron1971templeforum}. Mobile audio tours and electronic guidebooks show that in-situ guidance can reshape pacing, attention, social interaction, and participation roles~\cite{aoki2002sottovoce,grinter2002revisiting,woodruff2001guidebooks,goffman1981formsoftalk}. Outdoor AR extends these questions from exhibits to movement through place: the Touring Machine established wearable, location-aware urban exploration~\cite{feiner1997touring}; ARLang attached learning labels to outdoor objects~\cite{caetano2023arlang}; and site-specific AR storytelling links spatial trajectories to narrative structure and authorial intent~\cite{shin2023howspace}. More recent touring systems likewise show that placement, spatial awareness, and environmental competition shape real-world use~\cite{matviienko2022arsightseeing,ghaemi2023placement}. Together, this work motivates examining how visitor-timed conversational interpretation can share work with human staff across a freely ordered public exhibition route.

A second line of work examines how guide representation and spatial behaviour shape experience. Comparisons of real, realistic, abstract, and audio-only guides show that visual representation affects co-presence, perceived humanness, and attention to content~\cite{rzayev2019presence}; virtual-museum studies likewise find that stylised humanoid and animated-object guides support different kinds of engagement and questioning~\cite{lopezgarcia2024speakingobjects,schott2024guideapprentices}. Other XR research emphasises proxemics, motion, non-verbal cues, spatialised audio, and referential behaviour~\cite{wang2019agents,huang2022proxemics,congdon2023mixedagents,liu2026agenthands,cheng2026auditoryagents,kaschub2026referentialassistant}. Together, these studies show that embodiment can cue role and attention across varied guide forms. They primarily concern indoor or wholly virtual settings; our study asks how a deliberately artificial, non-humanoid guide remains legible while visitors walk through an outdoor MR exhibition.

Finally, conversational agents have been explored for guidance in cultural venues, from embodied docents to museum chatbots and LLM-supported interpretation~\cite{kopp2005museumguide,trichopoulos2023chatgpt4guide,chen2025museumguideprefs,trichopoulos2023llmheritage,su2025simviews,ho2025visualagents}. Recent XR systems investigate accessible guides, conversational task assistance, adaptive invocation, multimodal guidance, and proactive facilitation~\cite{collins2026llmvrguide,madritsch2026arconversation,wells2025llmxr,cascarano2025seehear,wang2026aispeakfirst}. Navigation Pixie further shows that an on-demand agent can change dwell time and free exploration in commercial virtual environments~\cite{yanagawa2025navigationpixie}. This literature establishes invocation timing, grounding, embodiment, and user control as design concerns. It leaves open how assigning a conversational agent as the default source of artwork interpretation changes explanation access, spatial attention, visitor self-curation, and human responsibility in a public, safety-constrained outdoor MR setting.

\section{Dream-Butterfly System and Guidance Design}
\label{sec:system}

Dream-Butterfly was deployed as a situated interpretation layer for a public outdoor MR exhibition. The design responds to three field needs: artwork-specific questions should be possible at self-chosen moments; the guide should remain spatially legible while visitors walk, turn, and manage outdoor safety; and human staff should remain clearly responsible for safety, wayfinding, device support, and contingencies.

At the infrastructure level, the field implementation ran on \textbf{Pico 4U} head-mounted displays~\cite{pico4ultraOfficial}. To register the campus-scale exhibition, we used an \textbf{Insta360} scan of the roughly 26,000~m$^2$ university campus outdoor site~\cite{insta360Official} to build \textbf{simultaneous localisation and mapping (SLAM)}-based scene features~\cite{macario2022comprehensive}, combined with GPS-supported coarse placement for anchoring virtual artworks and labels in the open environment. This registration layer defined the MR exhibition runtime: each registered artwork and label was assigned an artwork ID; the HMD handled localisation, spatial anchoring, interaction state, subtitle display, and the butterfly's world-space motion; and a local middleware server orchestrated retrieval and calls to the deployed automatic speech recognition (ASR), large language model (LLM), and text-to-speech (TTS) services.

\subsection{Guidance Ecology and System Flow}

\begin{figure}[t]
  \centering
  \includegraphics[
    width=0.80\linewidth,
    height=0.45\textheight,
    keepaspectratio
  ]{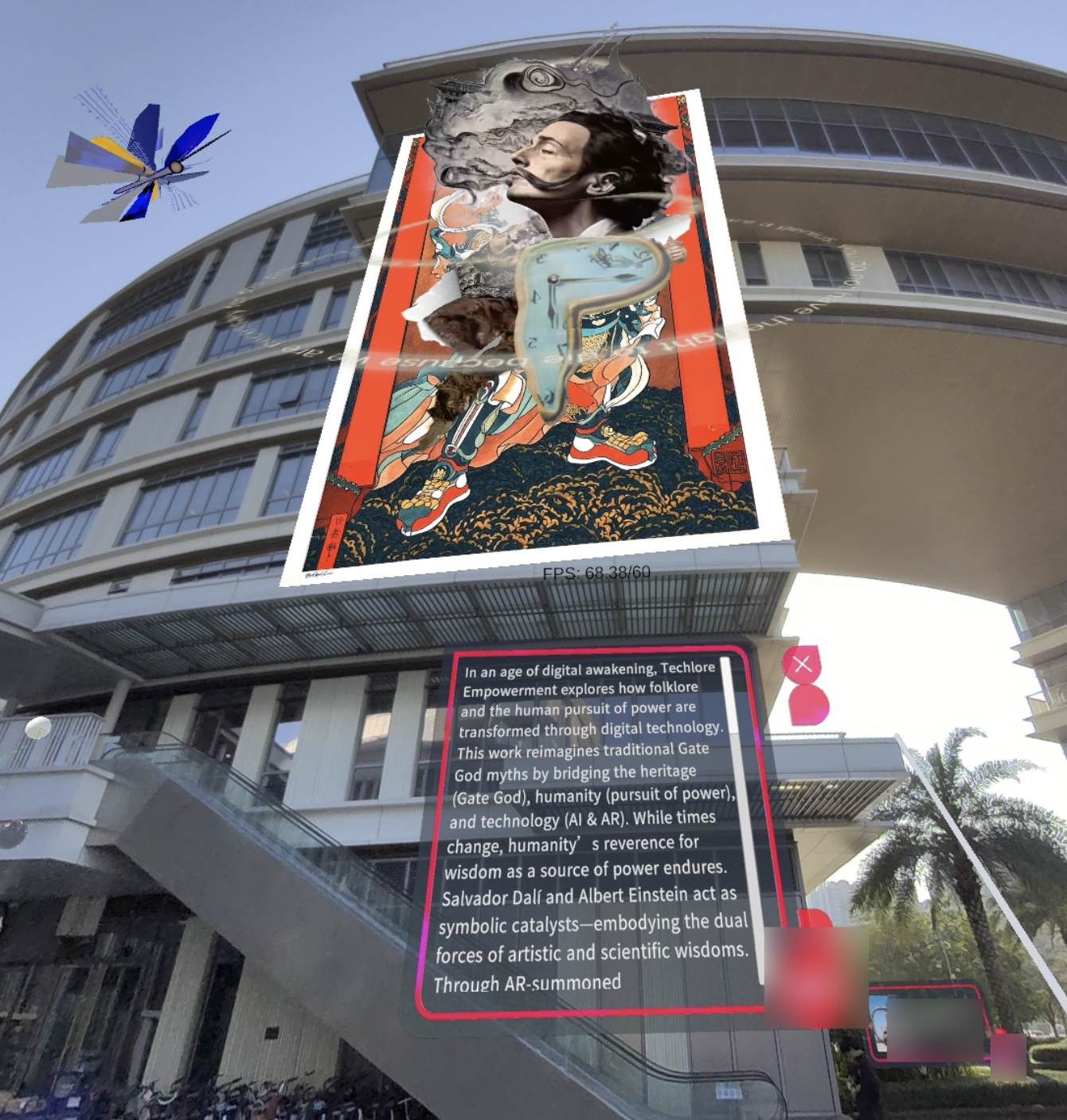}
  \caption{A representative in-situ MR label used as first-pass artwork context. Labels support lightweight access to registered work information, but their limited display space and non-conversational format constrain timely clarification during roaming.}
  \Description{Headset view in an outdoor mixed reality exhibition. A virtual artwork is anchored on a building facade. A semi-transparent information panel displays text about the artwork.}
  \label{fig:libai}
\end{figure}

The exhibition combined four guidance channels: human staff for route-level support, safety monitoring, device troubleshooting, and contingencies; virtual labels registered to individual artworks for first-pass information; Dream-Butterfly for summonable conversational interpretation; and visitors themselves for pacing and route-level choices. Static labels were useful for lightweight context but constrained deeper clarification because of limited display area and non-conversational presentation (Fig.~\ref{fig:libai}). Dream-Butterfly extended this ecology with summonable, artwork-specific dialogue.

In the HMD, Dream-Butterfly remains near the visitor in an idle following state. Visitors point the controller ray at a registered artwork or label to set that work as the active context, then press and hold the controller grip to summon the guide. The butterfly returns towards the hand and lands nearby as a readiness cue; a microphone indicator confirms that the system is listening; and the visitor asks a spoken question. Each dialogue turn triggers speech transcription, retrieval over exhibition materials, response generation, and speech rendering with optional subtitles.

Artwork identity is provided by the MR exhibition runtime: each spatially anchored artwork and label has a unique artwork ID, and the runtime exposes the active ID from the selected label or encountered work context. Dream-Butterfly uses this ID to scope retrieval before querying the dialogue backend, so questions are answered using materials for that artwork by default. This closed-world scope covers registered exhibition works. If the active ID is unavailable or ambiguous, visitors are directed to select a label, move closer to the intended artwork, or ask staff.

\subsection{Embodiment and Spatial Legibility}

We use a lightweight non-humanoid embodiment. Prior work shows that guide appearance changes co-presence, perceived humanness, attention, and the kinds of questions visitors ask~\cite{rzayev2019presence,lopezgarcia2024speakingobjects}. The butterfly form provides a small, socially addressable presence whose visible artificiality supports a clearly bounded guide role.

The final asset uses an abstracted butterfly silhouette, semi-transparent wings, and a high-contrast yellow--blue palette for outdoor readability. The form signals an artificial guide while staying light enough for peripheral awareness during walking; the full asset pipeline is provided in the supplementary material.

\subsection{Explicit Invocation and Visible Re-entry}

Dream-Butterfly uses user-initiated turn-taking. Press-and-hold grip summoning provides an explicit intent signal, while the return-to-hand motion and microphone indicator make the start of a dialogue turn visible. Visitors can stop the dialogue turn through the interface and resume roaming.

We use \emph{spatial re-entry} to describe the guide's visible world-space return to a hand-relative dialogue position after drifting out of view during roaming.

World-space following keeps the butterfly findable during head and body motion while preserving its situated character. In idle flight, the butterfly pursues a moving world-space goal point computed from the visitor's camera transform and an authored follow offset. During summoning, the goal switches to a hand-relative landing offset. We apply filtering to smooth the following trajectory~\cite{pan2023flow}. A small hover modulation keeps the agent visibly alive, and wing-flap speed changes communicate acceleration, cruising, and settling.

This behaviour intentionally allows brief lag during rapid head or body motion. The resulting drift-and-recovery movement gives the butterfly the character of a situated companion keeping up with the visitor. The summon motion also serves as a spatial re-entry cue: when the guide has moved out of view, calling it brings it back through a visible movement path into the display.

\subsection{Runtime-Scoped Retrieval and Multilingual Response}

Dream-Butterfly uses retrieval-augmented generation over exhibition materials to keep responses tied to curator and artist intent~\cite{lewis2020rag}. The knowledge base and docent preparation used the same label-level corpus, comprising artwork descriptions and curator- or artist-provided text published on the exhibition labels. Because source materials were multilingual and unevenly distributed, we treat the creator-provided language version as the canonical evidence source for each artwork and use team-produced translations as auxiliary materials for accessibility.

At runtime, the visitor's query is transcribed and paired with the active artwork ID from the MR runtime. When needed, the query is translated into the canonical language for that artwork before retrieval. Retrieval is first scoped to canonical materials for the current artwork; supplementary multilingual materials are used only when canonical evidence is sparse. Retrieved passages and source metadata are passed to the generator, which is instructed to prioritise evidence, avoid unsupported claims, and defer or ask for clarification when evidence is insufficient. The response is rendered in the visitor's selected language through speech and optional subtitles using deployed ASR/LLM/TTS services~\cite{sun2024hunyuan}.

This conservative grounding strategy scopes Dream-Butterfly to the exhibition and, by default, to the currently encountered artwork.

\subsection{Onboarding and Boundary Cues}

\begin{figure}[t]
  \centering
  \includegraphics[width=\linewidth]{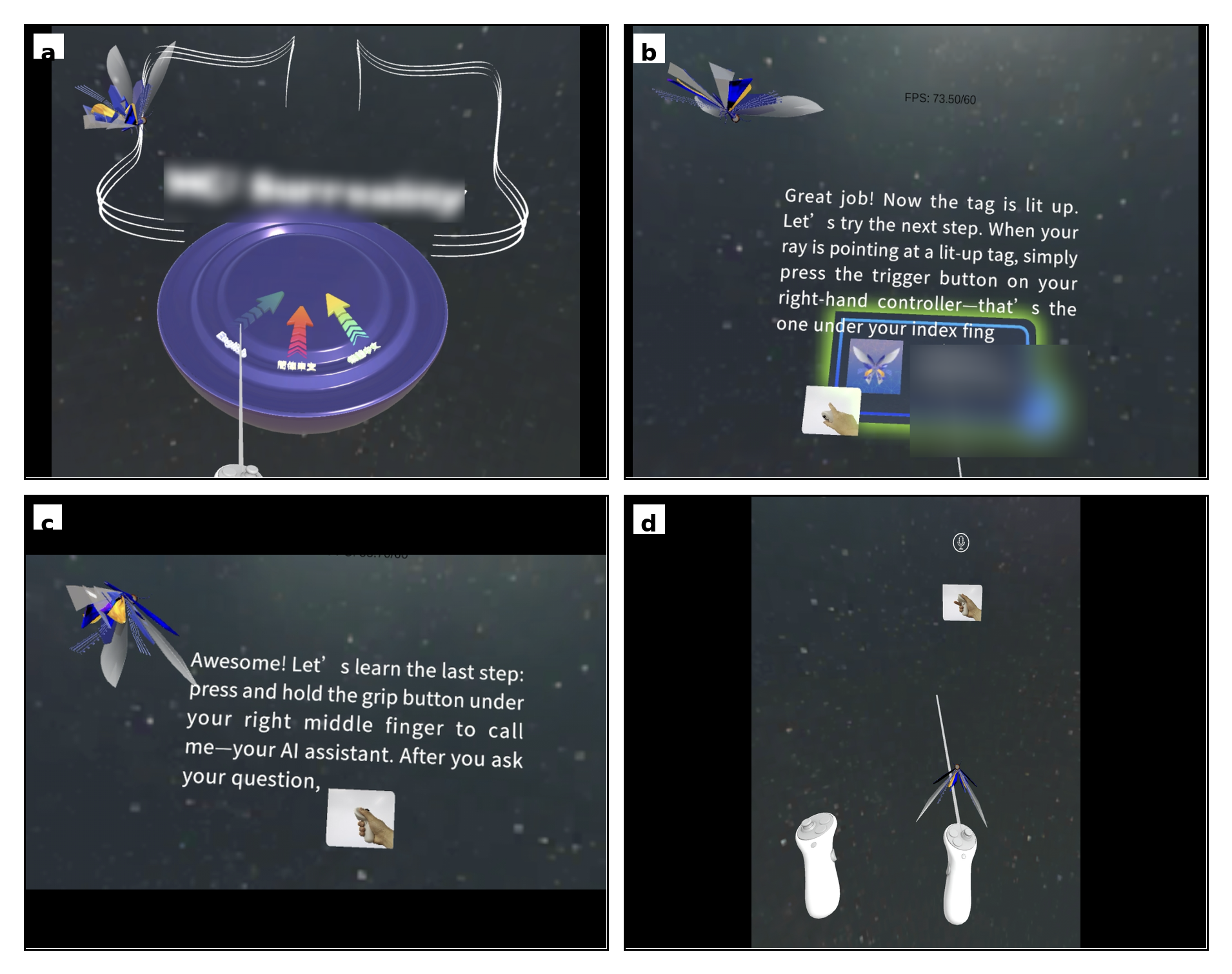}
  \caption{Shared in-headset onboarding tutorial for the interaction mechanics used in both conditions: language selection, label access, artwork selection, summoning, and microphone state. A separate condition-specific briefing established the default source of artwork interpretation.}
  \Description{A four-panel onboarding figure showing language selection, label interaction, ray-based artwork selection, summoning the butterfly guide with a press-and-hold grip gesture, and an active microphone state during dialogue.}
  \label{fig:onboarding-ui}
\end{figure}

Before entering the exhibition route, visitors completed a shared in-headset onboarding tutorial covering language selection, virtual label reading, Dream-Butterfly summoning, and microphone/listening state (Fig.~\ref{fig:onboarding-ui}). The shared tutorial covered system operation and made clear that human docents remained available for non-artwork questions and support. A separate condition-specific briefing, described in Section~\ref{sec:study-design}, established whether Dream-Butterfly or the human docent served as the default source of artwork interpretation.

Together, these mechanisms make Dream-Butterfly a summonable, spatially situated, runtime-grounded interpretation layer. The next section evaluates how visitors experienced this design, and how making it the default interpretive channel reshaped explanation access, spatial attention, and responsibility handoffs compared with docent-default guidance.

\section{In-the-Wild Field Evaluation}
\label{sec:study}

We evaluated Dream-Butterfly during a three-month public outdoor MR exhibition through a comparison of two guidance conditions. In both conditions, human staff remained present for safety, wayfinding, device support, and contingencies; the default source of artwork interpretation varied.

\subsection{Participants and Setting}

We used on-site intercept recruitment during the public exhibition. Visitors who were about to begin the exhibition route were invited to join an optional study session. Participants who consented experienced the exhibition under one of the two guidance conditions, then completed a post-visit questionnaire and a semi-structured interview.

We recruited 24 participants in total, with 12 participants per condition. The sample included 12 men, 11 women, and 1 non-binary participant, aged 16--66 years. Participants selected Mandarin ($n=13$; agent-default: 6, docent-default: 7), English ($n=7$; agent-default: 4, docent-default: 3), or Cantonese ($n=4$; agent-default: 2, docent-default: 2) as the guiding language during the experience. Participant IDs B1--B12 and G1--G12 are retained only as anonymised source identifiers. Participants varied in prior XR experience, from almost none to extensive experience; a participant-level demographic table is provided in the supplementary material.

\subsection{Study Design and Conditions}
\label{sec:study-design}

We used a between-subject field comparison with two guidance conditions. The labels \emph{agent-default} and \emph{docent-default} indicate the first resort for artwork interpretation, while both human and system channels were present. Assignment followed staff availability and exhibition operations, yielding a non-random field comparison. Both conditions used the same exhibition area, HMD setup, safety briefing, and onboarding tutorial, and sessions were scheduled within comparable time windows and route segments when feasible.

In the \textbf{agent-default condition}, Dream-Butterfly was introduced as the default channel for artwork interpretation. Participants were explicitly told that the accompanying docent remained available for any non-artwork need, including route choices, safety or comfort concerns, and technical problems. Docents avoided artwork interpretation except when needed to resolve a breakdown or keep the visit coherent.

In the \textbf{docent-default condition}, human docents provided the primary interpretive narration and answered artwork-related questions during the visit. Dream-Butterfly was still introduced during onboarding and remained available as optional support, reflecting the realistic mixed-channel nature of public MR exhibitions. Post-visit questionnaires and interviews documented how participants distributed guidance across docents, the agent, labels, and themselves.

Four docents covered the sessions: three guided in Mandarin and English, and one handled Cantonese sessions. Every docent completed two rounds of training covering all participating artists and the curatorial material published on the exhibition labels. They were expected to introduce each work's artist, motivation, and main idea. Their preparation used, and was bounded to, the same label-level corpus stored in Dream-Butterfly's retrieval-augmented generation (RAG) system. Further condition and training details are provided in the supplementary material.

\subsection{Procedure}

All participants first received a safety briefing and basic HMD operation instructions. They then completed the same in-headset onboarding tutorial, covering language selection, virtual label access, and Dream-Butterfly summoning. Participants experienced the outdoor MR exhibition under one of the two conditions. After the visit, they completed a questionnaire and took part in a 20--30 minute semi-structured interview.

\subsection{Measures}

The post-visit questionnaire included three groups of measures. First, custom 7-point Likert items captured constructs aligned with our research questions: \textit{Explanation Access \& Interaction} (4 items), \textit{Immersion \& Engagement} (4 items), and \textit{Role Clarity \& Handoff} (5 items). Second, a two-item workload probe adapted from NASA-TLX measured \textit{Effort} and \textit{Frustration}~\cite{hart1988nasatlx}, and UEQ-S captured pragmatic and hedonic experience quality~\cite{schrepp2017ueqs}. Third, a \textit{Responsibility Distribution Matrix} asked participants to allocate 100 points across Human staff, Dream-Butterfly, and \textit{Virtual labels / Myself \& companions} for five domains: navigation, explanation, pacing, troubleshooting, and safety.

Participants in the agent-default condition additionally completed the System Usability Scale (SUS) for Dream-Butterfly~\cite{brooke1996sus}. Unless otherwise noted, custom items used a 7-point scale from 1 (Strongly disagree) to 7 (Strongly agree). Reverse-worded items were reverse-scored before analysis, and composite scores were computed as the mean of constituent items. The custom composites serve as field-specific indicators aligned with our research questions. Full questionnaire items are provided in the supplementary material.

Internal consistency was $\alpha=0.70$ for \textit{Explanation Access \& Interaction} and $\alpha=0.64$ for \textit{Immersion \& Engagement}. The role-clarity items are reported individually.

\subsection{Interviews and Analysis}

The semi-structured interviews focused on explanation access, successful or failed guidance moments, responsibility distribution among staff, Dream-Butterfly, labels, and the visitor, and breakdowns related to language, safety, device use, or public interaction. Interviews were conducted in Mandarin or English; participants who used Cantonese guidance were interviewed in Mandarin. All Cantonese-guided participants were comfortable being interviewed in Mandarin, and language-specific breakdowns were explicitly probed. With consent, interviews were recorded, transcribed, de-identified, and analysed in the original language where possible.

For quantitative analysis, we report descriptive statistics and use Mann--Whitney U tests for between-condition comparisons because of the small sample size and ordinal response scales. Associations between questionnaire constructs are reported as exploratory Pearson correlations.

For qualitative analysis, we used reflexive thematic analysis~\cite{braun2019reflecting}. We reread the post-visit interview transcripts for familiarisation, then coded explicit accounts of explanation timing, attention shifts, role decisions, breakdowns, and recovery. Codes were compared within and across conditions, grouped into candidate themes, and revised against discrepant as well as supporting cases. Research-team discussions challenged interpretations and refined theme boundaries. Quotations originally spoken in Chinese were translated into English by the first author for reporting. The qualitative analysis drew on post-visit accounts.

\subsection{Ethics}

The study protocol was reviewed and approved by an institutional ethics review board, including procedures for interviewing minor participants. For minor participants, we followed the approved assent/guardian-consent procedure. Participation was voluntary, and visitors could decline or withdraw without affecting exhibition access or on-site support. Participants received 50 RMB as compensation. Questionnaire responses and interview transcripts were de-identified and linked using anonymous participant IDs. Because the study took place in an open outdoor setting, staff could intervene at any time for safety, comfort, or device-related issues.

\section{Findings}
\label{sec:findings}

Dream-Butterfly was available in both conditions. The labels \emph{agent-default} and \emph{docent-default} identify the channel assigned primary responsibility for artwork interpretation.

\subsection{Quantitative Results}
\label{subsec:findings-quant}

Given the small sample and ordinal responses, between-condition comparisons use two-sided asymptotic Mann--Whitney U tests with tie and continuity correction; associations are reported as exploratory Pearson correlations. We report $U$ for the agent-default group and effect size $r=|z|/\sqrt{N}$. Complete medians, group sample sizes, $U$, $p$, and $r$ values for all tested outcomes are provided in the supplementary quantitative-results table.

\begin{figure*}[t]
  \centering
  \includegraphics[width=\textwidth]{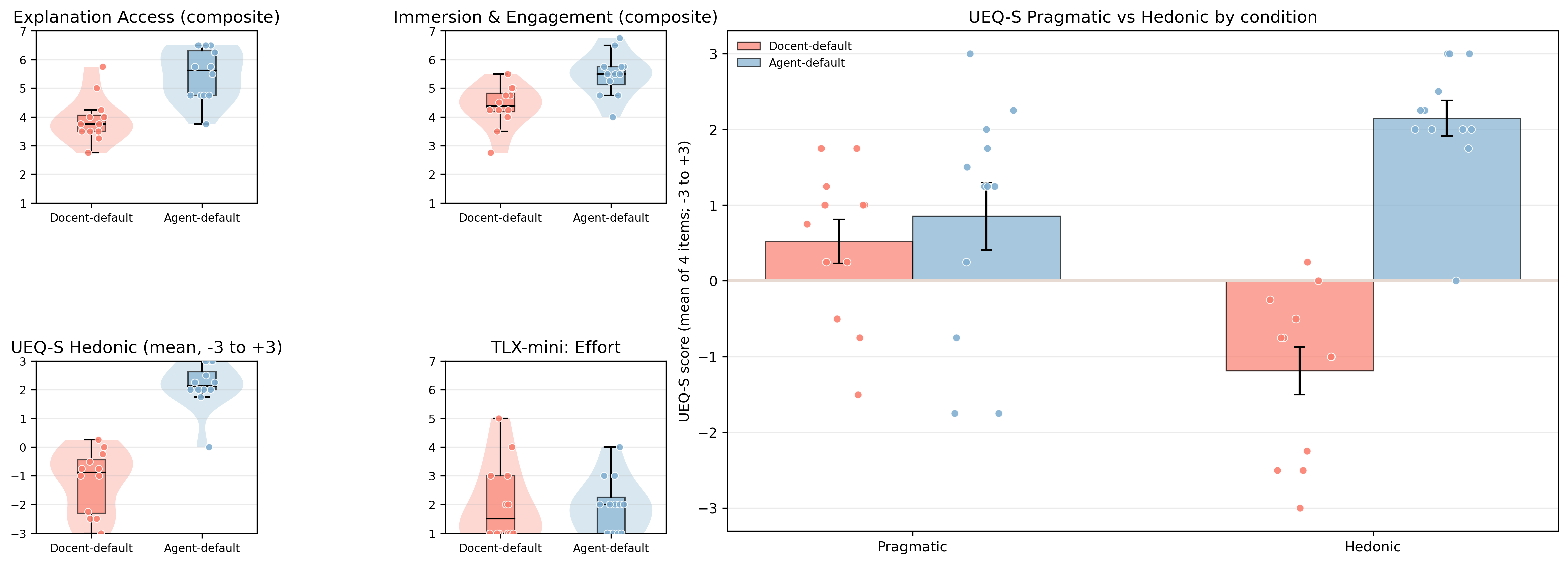}
  \caption{Questionnaire outcomes by condition. \textbf{Left:} Distributional comparisons for interpretive access, immersion/engagement, hedonic quality, and effort. \textbf{Right:} UEQ-S pragmatic and hedonic quality scores by condition. Bars show means $\pm$ one standard error; points represent individual participants.}
  \Description{Two-panel figure comparing docent-default and agent-default conditions. Left panel shows distribution plots for Explanation Access, Immersion and Engagement, UEQ-S Hedonic quality, and TLX-mini Effort. Access, immersion, and hedonic quality are higher in agent-default, while effort remains similarly low. Right panel shows UEQ-S Pragmatic and Hedonic quality scores; pragmatic scores overlap, while hedonic scores are higher for agent-default. Bars show means with one standard error above and below each mean; points represent individual participants.}
  \label{fig:quant-outcomes}
\end{figure*}

\paragraph{RQ1. Pull-based interpretation was associated with stronger explanation access and immersion.}
When Dream-Butterfly served as the default way to access explanations, participants could \emph{pull} interpretation at moments of curiosity and at their own pace.
This shift appears in higher access and immersion ratings (Fig.~\ref{fig:quant-outcomes}).
In the agent-default condition, \textit{Explanation Access \& Interaction} was higher ($M=5.46$, $Mdn=5.63$, $n=12$) than in docent-default ($M=3.92$, $Mdn=3.75$, $n=12$; $U=128.0$, $p=0.0013$, $r=0.66$).
\textit{Immersion \& Engagement} showed a similar difference (agent-default: $M=5.48$, $Mdn=5.50$, $n=12$; docent-default: $M=4.38$, $Mdn=4.38$, $n=12$; $U=124.0$, $p=0.0028$, $r=0.61$).
This pattern suggests that on-demand conversational interpretation can align explanation with visitors' situated attention during outdoor roaming.

We detected no statistically significant between-condition differences in workload. TLX-mini \textit{Effort} (agent-default: $Mdn=2.00$, $n=12$; docent-default: $Mdn=1.50$, $n=12$; $U=75.5$, $p=0.8551$, $r=0.04$) and \textit{Frustration} (agent-default: $Mdn=2.00$, $n=12$; docent-default: $Mdn=1.50$, $n=12$; $U=78.0$, $p=0.7306$, $r=0.07$) showed small observed effects; these tests do not establish equivalence.
UEQ-S clarifies the character of the improvement.
The agent-default condition was associated with higher \textit{Hedonic Quality} ($Mdn=2.13$ vs.\ $-0.88$; $n=12$ per group; $U=142.5$, $p=0.000049$, $r=0.83$), while \textit{Pragmatic Quality} showed no statistically significant difference ($Mdn=1.25$ vs.\ $0.88$; $n=12$ per group; $U=91.0$, $p=0.2830$, $r=0.22$; Fig.~\ref{fig:quant-outcomes}).
SUS for Dream-Butterfly in the agent-default condition averaged $65.6$ (range $40$--$85$), suggesting moderate usability with clear room for refinement in a field deployment.

\begin{figure*}[t]
  \centering
  \includegraphics[width=\textwidth]{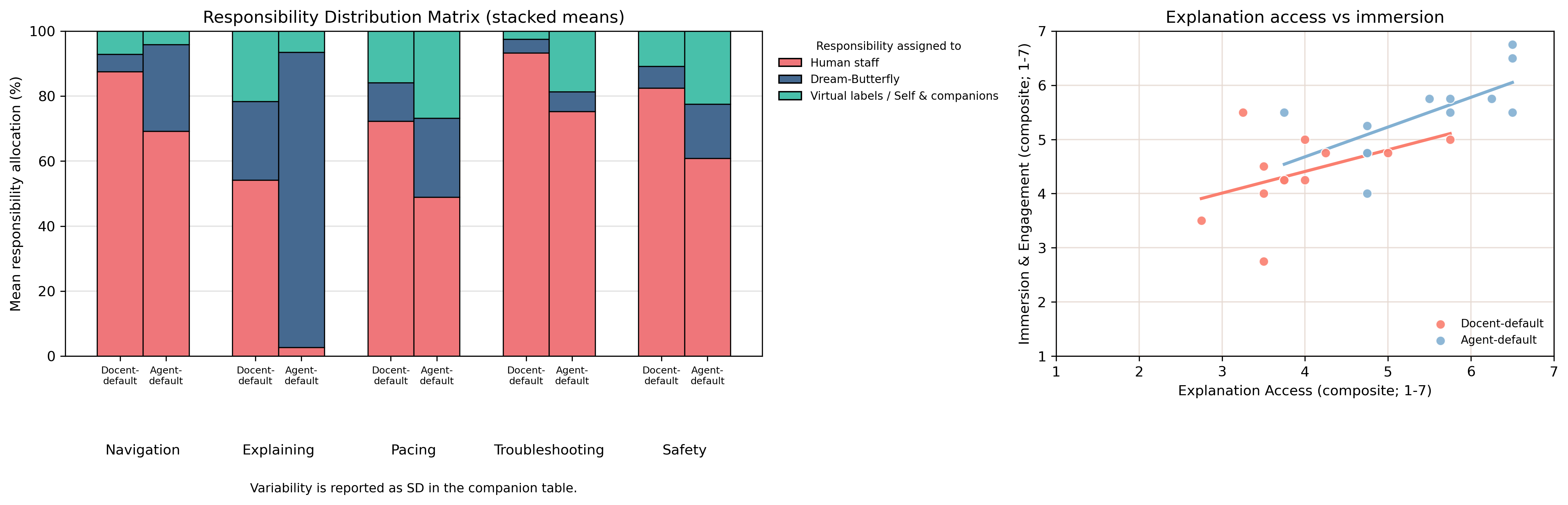}
  \caption{Role ecology by condition. \textbf{Left:} Mean responsibility allocations across five guiding domains; SDs are reported in the supplementary table. \textbf{Right:} Explanation Access and Immersion/Engagement with per-condition trend lines; the reported correlation is computed across all participants.}
  \Description{Two-panel figure. Left panel is a 100 per cent stacked bar chart across Navigation, Explaining, Pacing, Troubleshooting, and Safety. Each domain compares docent-default and agent-default conditions, segmented into Human staff, Dream-Butterfly, and Virtual labels or Self and companions. Human staff dominate Troubleshooting and Safety in both conditions, while Dream-Butterfly dominates Explaining in agent-default. Right panel is a scatter plot showing a positive association between Explanation Access and Immersion and Engagement.}
  \label{fig:responsibility-corr}
\end{figure*}

Across all participants, easier explanation access also tended to coincide with higher immersion ($r=0.74$, $p<0.001$; Fig.~\ref{fig:responsibility-corr}, right).
This covariation is consistent with a timing account in which situated, on-demand interpretation supports engagement.

\paragraph{RQ2. Role boundaries became clearer and visitors inherited more self-curation work.}
Positioning Dream-Butterfly as the default interpretation channel made the guidance ecology easier to read while foregrounding visitors' own role in shaping the visit (Fig.~\ref{fig:responsibility-corr}).
Agent-default participants more strongly agreed that they understood what the primary guide could help with ($Mdn=6.00$ vs.\ $5.00$; $U=133.5$, $p=0.00022$, $r=0.75$), found it easier to decide whom to turn to first when something happened ($Mdn=6.00$ vs.\ $4.00$; $U=137.5$, $p=0.00011$, $r=0.79$), and perceived switching among staff, the agent, and themselves as smoother ($Mdn=6.00$ vs.\ $4.00$; $U=116.0$, $p=0.0096$, $r=0.53$); each comparison included $n=12$ per group.
These items indicate situated role legibility: in outdoor MR, visitors continually face small uncertainties, and the default-role assignment shaped how quickly they could identify the right kind of help.

At the same time, the item about unexpected responsibility moved in the opposite direction.
After reverse-coding (higher = \emph{less} burden), the agent-default condition scored lower than docent-default ($Mdn=2.00$ vs.\ $4.00$; $n=12$ per group; $U=34.0$, $p=0.0281$, $r=0.45$).
This tension is central to our findings: making interpretation more available and visitor-timed also shifted more micro-decisions about pacing, attention, and when to ask questions onto visitors.

The Responsibility Distribution Matrix makes this division of labour explicit (Fig.~\ref{fig:responsibility-corr}, left).
In both conditions, staff remained the dominant safety net for troubleshooting and safety.
For explaining content, responsibility shifted strongly towards Dream-Butterfly in the agent-default condition ($90.8\%$, $SD=7.5$), whereas docent-default remained a mixed ecology (staff: $54.2\%$, $SD=21.5$; Dream-Butterfly: $24.2\%$, $SD=17.3$; Virtual labels/Self: $21.7\%$, $SD=12.7$).
Means and SDs for every party, domain, and condition are reported in the supplementary responsibility-allocation table. The allocations show that agent-default sharpened interpretive role boundaries and made visitors' own management of the experience more salient.

\subsection{Qualitative Mechanisms}
\label{subsec:findings-qual}

Interview themes help interpret how experience and responsibility differed across the default-role arrangements.
We focus on three mechanisms: timing and granularity of interpretation, attention regulation during roaming, and responsibility handoffs among staff, Dream-Butterfly, and visitors.

\subsubsection{Timing and granularity: why visitors still summoned Dream-Butterfly}
Interpretation repeatedly emerged as a timing-sensitive need.
Questions often appeared while visitors watched dynamic or abstract works unfold, and sometimes required work-specific details that were hard to provide at group-tour pace.
Front-end and back-end communication records confirmed that \textbf{9 of 12 docent-default participants} invoked Dream-Butterfly at least once. The count comes from system records, and the explanations below come from post-visit interviews.

Docent-default participants described using the butterfly when narration and curiosity were temporally misaligned.
G5 recalled that when an animated work changed in a confusing way, the docent was still explaining the previous work to a companion: ``I summoned the butterfly right away'' (G5).
They also used Dream-Butterfly to resolve granularity mismatches, such as when a docent explained the exhibition's overall theme but visitors wanted details about a particular work.
In agent-default, participants described the same pattern as the normal mode of interpretation.
As B4 put it, ``The artworks are dynamic. Many require you to stop and watch how they change. When I had questions, the butterfly's explanation helped me understand'' (B4).
This helps explain why pull-based explanation felt more available: it matched the moment at which questions arose.

\subsubsection{Attention regulation: micro-pauses and breakdowns}
Participants described Dream-Butterfly as a low-friction presence during roaming.
Its idle flight stayed in peripheral awareness, and the visible summon-to-hand motion created continuity at the start of dialogue.
B1 described briefly losing sight of the guide and then seeing it return overhead, reinforcing its character as a situated guide moving with the visitor.

Explanation channels also shaped attention between virtual artworks and the physical-social setting.
In docent-default, G7 appreciated the docent but felt narration could pull them out of the MR experience: ``I respect the docent's explanations, but it always makes me feel pulled back into reality'' (G7).
In agent-default, dialogue often functioned as a visitor-timed micro-pause: participants asked when ready, received a short explanation, and then resumed viewing.

This attention benefit depended on public comfort and language fidelity.
Public speech through an HMD could make visitors self-conscious. B3 noted, ``Wearing an HMD outside already draws attention. Talking to the butterfly feels even weirder. I'd rather stay close to the docent so it looks like I'm attending an exhibition and people won't stare at me'' (B3). Language instability could also break presence.
B6 described the Cantonese output as mixing Cantonese, Mandarin, and English, which was ``very immersion-breaking'' (B6).
Others reported occasional off-topic or template-like answers, after which they stopped the dialogue and returned attention to the surrounding exhibition.
The agent supported attention when timing and language felt situated; publicness and language fidelity remained important failure points.

\subsubsection{Responsibility handoffs: staff safety, visitor pacing}
Across both conditions, participants treated staff as the accountable layer for safety and device contingencies.
Outdoor MR could blur virtual and physical boundaries around stairs, water, traffic, and device failures, making human oversight non-delegable. G7 explained, ``This MR exhibition makes it hard to distinguish virtual from real, especially with the lake (and works on the water), and sometimes vehicles passing by'' (G7).
When technical problems occurred, participants generally handed them back to staff even if they briefly tried asking Dream-Butterfly.

Where the conditions diverged was pacing and inquiry.
Several docent-default participants described following the docent's route and tempo, while agent-default participants more often described detours, corner exploration, and self-paced stopping.
One agent-default participant tried asking Dream-Butterfly about a campus sculpture outside the exhibition map, showing how free roaming could exceed its closed-world scope.
This freedom encouraged exploration and made visitors responsible for deciding what counted as worth asking.

Human staff provided proactive forms of care, including noticing dizziness, adjusting setup, and encouraging gradual adaptation. Proactive prompting, comfort monitoring, and early confidence-building remained human work. Agent-default guidance made interpretation more pullable and required visitors to signal discomfort, initiate questions, and translate diffuse curiosity into askable prompts.

\subsection{Summary of Findings}
For \textbf{RQ1}, agent-default participants reported more available, visitor-timed explanations, higher immersion/engagement, and stronger hedonic quality; no statistically significant differences were detected in effort or frustration. For \textbf{RQ2}, concentrating interpretation in Dream-Butterfly clarified the guidance ecology: staff retained safety and contingency responsibilities, while visitors assumed more timing and inquiry work. Embodied conversational guidance therefore distributes roles across human and system channels.

\section{Discussion}

\subsection{The default interpretive channel as a role assignment}

Our findings concern a default-role assignment across a mixed guidance ecology. Both conditions had access to Dream-Butterfly; the first resort for artwork interpretation varied. In the agent-default condition, staff focused on safety, wayfinding, and device support while Dream-Butterfly carried primary interpretive work. In the docent-default condition, staff provided primary narration while Dream-Butterfly remained available as optional support. This framing matters because outdoor MR exhibitions combine staff, systems, labels, companions, and self-curation in public, mobile settings~\cite{Djuric2025PublicSpaceArtMRAR,Trefzger:2025:ymt,Moon2024CrossDeviceMRArt}.

A \emph{role contract} captures who visitors were expected to ask first and which responsibilities remained visibly human. This aligns with conditional delegation, where people hand off subtasks based on timing, uncertainty, and perceived competence~\cite{Lai2022ConditionalDelegation}, and with work on making AI role boundaries explicit~\cite{10.1145/3719160.3736629}. Optional use in docent-default illustrates how visitors combined channels within the mixed guidance ecology. For XR system evaluation, the reusable object of study is the design that combines invocation, visible re-entry, artwork-specific grounding, and human role boundaries.

\subsection{Pull-based interpretation supports attention and redistributes self-curation work}

Timing appears central to the experiential benefit of agent-default guidance. Pull-based dialogue acted as a visitor-timed ``micro-pause'' that aligned explanation with situated curiosity: visitors could stop, ask, receive a short interpretation, and return attention to the artwork. Prior work on an on-demand navigation agent likewise found changes in dwell time and free exploration~\cite{yanagawa2025navigationpixie}; our field results connect that user-initiated timing to artwork explanation and responsibility sharing. This is particularly relevant outdoors, where walking, crowds, wayfinding, and safety checks fragment attention, and externally paced narration can feel like being pulled back into the physical-social world~\cite{Djuric2025PublicSpaceArtMRAR}. Public MR also creates a visibility problem: head-worn interaction can isolate the wearer from co-located audiences or make speaking to an agent socially exposed~\cite{Trefzger:2025:ymt,Moon2024CrossDeviceMRArt}. Publicness remained visible, while summonable, short-burst interaction made interpretation spatially re-acquirable and interruptible.

The same default-role assignment also explains the self-curation cost. When interpretation becomes pullable, visitors inherit more decisions about when to pause, what to ask, how much depth to pursue, and when to move on. This redistribution of guiding labour creates a need for inquiry scaffolds, lightweight prompts, and graceful handoffs that help visitors formulate questions while walking, staying oriented, and managing public exposure.

\subsection{Language fidelity must be quality-gated when the agent is the default interpretive channel}

The Cantonese breakdowns show that language fidelity directly affects presence in conversational MR. Participants experienced mixed-language output as inaccurate, immersion-breaking, and damaging to credibility. Accent and dialect cues shape credibility, alignment, and social presence~\cite{Wang2025DialectOlderAdultsAlignment}, while speech systems show uneven performance across racialised, regional, and global accent varieties~\cite{Koenecke2020RacialDisparitiesASR,DiChristofano2022AccentASRDisparities,Graham2024WhisperAccents,Michel2025NotRepresentation}. Recent Cantonese and Sinitic-variety resources suggest ecosystem progress~\cite{li2025wenetspeech,Wang2026WenetSpeechWu}, and public MR deployment requires end-to-end quality gating across speech recognition, response generation, and speech synthesis. When a language or dialect option is offered, especially for a default interpretive channel, conservative boundaries, fallback paths, and human handoff form part of deployment quality.

\subsection{Implications and Limitations}

We draw four design implications for embodied conversational guidance in public outdoor MR. First, spatial re-entry should be treated as a first-class interaction state: explicit input should make a drifting guide visibly re-acquirable through motion. Second, role boundaries should be explicit and redundant so visitors know whether to ask staff, the embodied guide, a label, or themselves for interpretation, safety, wayfinding, or troubleshooting. Third, agent-default designs should reduce self-curation overhead with inquiry scaffolds, suggested follow-up questions, and handoff cues that keep pull-based access supported. Fourth, language support should be quality-gated before an agent becomes the default interpretive channel, with conservative defaults and transparent fallbacks for unreliable coverage.

The small, field-constrained sample, non-random assignment, and field-specific custom measures limit causal inference and comparison of relative channel capability; the immersion composite also had modest internal consistency ($\alpha=0.64$). Despite a shared label-level corpus, access was asymmetric: agent-default docents withheld routine artwork interpretation, while Dream-Butterfly remained optional in docent-default, so a briefing-only effect under equivalent availability cannot be isolated. Guiding language was self-selected (Mandarin, $n=13$; English, $n=7$; Cantonese, $n=4$), with two Cantonese-guided participants per condition, precluding a condition-by-language test. The single exhibition site, one embodied-agent form, and its staff protocol further limit generalisation to other venues and guidance arrangements. The exploratory comparisons span multiple outcomes; individual $p$ values warrant caution.

No anonymised in-visit dialogue transcripts, dialogue-coding records, or synchronised utterance, route-position, and channel-choice timestamps were retained. System records verify invocation counts, while retrospective post-visit interviews provide explanations for channel choice; event-level timing and choice mechanisms remain outside the evidence base.

Novelty and repeat exposure were unmeasured. The animated butterfly and conversational interface may have increased initial summoning or hedonic ratings because they were new, and positive open-ended reactions cannot distinguish novelty from sustained value. Repeated-exposure or longitudinal deployments are needed to test whether the observed engagement persists after novelty fades. Future work should also test alternative role allocations and more robust language support.

\section{Conclusion}
In a campus-scale deployment with 24 visitors, agent-default guidance was associated with stronger explanation access, immersion/\allowbreak engagement, and hedonic quality than docent-default guidance and also shifted more timing and inquiry work to visitors. Staff retained safety, wayfinding, and device support in both conditions. We frame primary interpretation as a role assignment and derive implications for role boundaries, spatial re-entry, self-curation support, and language-quality gating in public MR.

\section{Disclosure about Use of LLM}
We used GPT-5.4 for Python code supporting data cleaning, analysis, and visualisation, and for light language editing. The authors made all analytic and interpretive decisions.

\begin{acks}
This work was supported in part by the National Key Research and Development Program of China under Grant 2024YFC3307602, and the Guangdong Provincial Talent Program, Grant No. 2023JC10X009.
\end{acks}

% =========================
% References
% =========================
\bibliographystyle{ACM-Reference-Format}
\bibliography{ref}
\label{MainPaperLastPage}

% The ACM/PCS main and independent supplement remain unchanged.
\clearpage
\onecolumn
\pagestyle{plain}

\input{appendix_body}

\end{document}

%% file: appendix_body.tex
\begin{center}
{\LARGE\bfseries Supplementary Material}\\[0.5em]
{\large Dream-Butterfly: Configuring Embodied Conversational Guidance for Public Outdoor Mixed Reality Exhibitions}\\[0.5em]
Dongyijie Primo Pan, Shuyue Li, Yawei Zhao, Junkun Long, Hao Li, and Pan Hui\\[0.5em]
{\small Corresponding author: Pan Hui; \nolinkurl{panhui@ust.hk}}\\[0.5em]
{\small Main article: \url{https://doi.org/10.1145/3822517.3848728}}
\end{center}

\noindent\textbf{Scope.}
This supplementary material extends the self-contained main paper with supporting details for the asset pipeline, participant demographics, condition briefing and docent preparation, study instruments, additional quantitative reporting, and motion equations.

\noindent\textbf{arXiv navigation.} In this arXiv version, the supplementary material referred to in the main paper is included below in the same PDF as Appendices A--F.

\appendix
\setcounter{figure}{0}
\setcounter{table}{0}
\setcounter{equation}{0}
\renewcommand{\thefigure}{S\arabic{figure}}
\renewcommand{\thetable}{S\arabic{table}}
\renewcommand{\theequation}{S\arabic{equation}}
\renewcommand{\theHfigure}{supp.\arabic{figure}}
\renewcommand{\theHtable}{supp.\arabic{table}}
\renewcommand{\theHequation}{supp.\arabic{equation}}

\section{Dream-Butterfly asset pipeline}
\label{app:asset-pipeline}

\begin{figure}[h]
  \centering
  \includegraphics[width=0.70\linewidth]{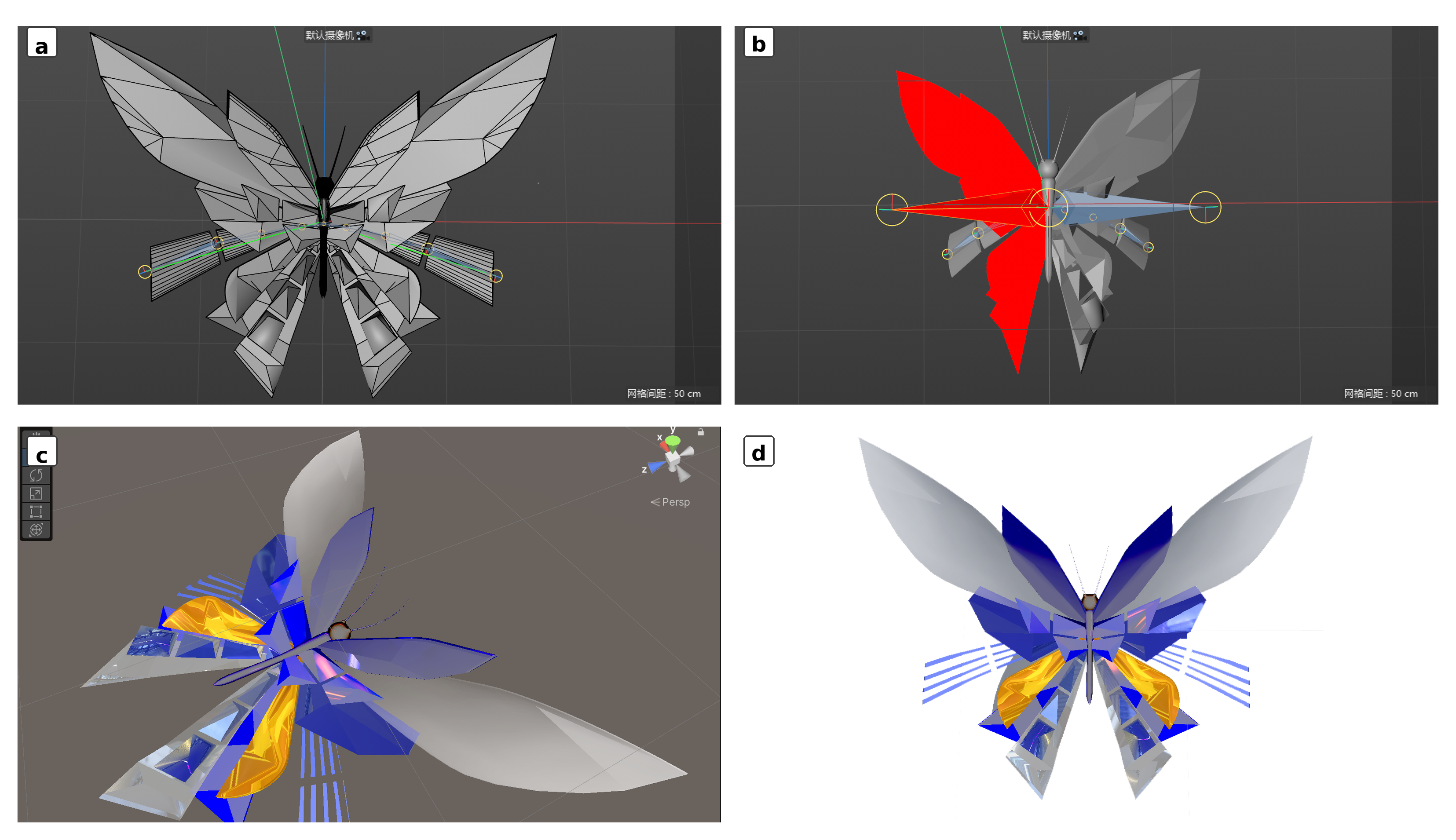}
  \caption{Dream-Butterfly asset pipeline: white-box geometry, rigging and skin weights, material/shader development, and final optimised asset.}
  \Description{A four-panel pipeline figure of the Dream-Butterfly guide asset. The panels show white-box geometry, rigging and skin weights, material and shader development, and the final optimised butterfly asset.}
\end{figure}

% -----------------------------
% Appendix A: Participant Demographics
% -----------------------------
\section{Participant demographics}
\label{app:demographics}

\renewcommand{\arraystretch}{1.15}
\small

\subsection{Participant demographics (agent-default condition; participant IDs B1--B12)}
\label{app:butterfly-demo}

% --- A方案：只对该表收紧 padding + 更保守的列宽 ---
\begingroup
\setlength{\tabcolsep}{3pt}      % 默认约6pt；收紧列间距，避免超出页边
\renewcommand{\arraystretch}{1.10}

\begin{longtable}{|p{0.06\linewidth}|p{0.09\linewidth}|p{0.05\linewidth}|p{0.13\linewidth}|p{0.10\linewidth}|p{0.37\linewidth}|p{0.10\linewidth}|}
\caption{Demographic profile of participants in the agent-default condition; B1--B12 are anonymised participant IDs.}
\label{tab:app-b-demo}\\
\hline
\textbf{ID} & \textbf{Gender} & \textbf{Age} & \textbf{Nationality} & \textbf{Guide lang.} & \textbf{Occupation} & \textbf{XR exp.} \\
\hline
\endfirsthead

\hline
\textbf{ID} & \textbf{Gender} & \textbf{Age} & \textbf{Nationality} & \textbf{Guide lang.} & \textbf{Occupation} & \textbf{XR exp.} \\
\hline
\endhead

B1  & Male       & 24 & Vietnamese & English   & Master's student                 & Almost none \\
\hline
B2  & Female     & 16 & Chinese    & Mandarin  & High school student              & Almost none \\
\hline
B3  & Non-binary & 22 & Chinese    & Cantonese & UG student                       & Some \\
\hline
B4  & Male       & 53 & Chinese    & Mandarin  & Electrical engineer              & Extensive \\
\hline
B5  & Female     & 53 & Chinese    & Mandarin  & Legal professional               & Almost none \\
\hline
B6  & Male       & 23 & Chinese    & Cantonese & Media producer                   & Extensive \\
\hline
B7  & Female     & 22 & Chinese    & Mandarin  & Film industry practitioner       & Some \\
\hline
B8  & Male       & 27 & French     & English   & Game developer                   & Very extensive \\
\hline
B9  & Female     & 52 & French     & English   & Administrative staff             & Some \\
\hline
B10 & Male       & 30 & Chinese    & Mandarin  & Education professional           & Extensive \\
\hline
B11 & Female     & 26 & Chinese    & Mandarin  & PhD student                      & Some \\
\hline
B12 & Male       & 40 & USA   & English   & Academic                         & Extensive \\
\hline

\end{longtable}
\endgroup

\subsection{Participant demographics (docent-default condition; participant IDs G1--G12)}
\label{app:guide-demo}

% --- A方案：同样处理第二张表，列规格统一，避免再次溢出 ---
\begingroup
\setlength{\tabcolsep}{3pt}
\renewcommand{\arraystretch}{1.10}

\begin{longtable}{|p{0.06\linewidth}|p{0.09\linewidth}|p{0.05\linewidth}|p{0.13\linewidth}|p{0.10\linewidth}|p{0.37\linewidth}|p{0.10\linewidth}|}
\caption{Demographic profile of participants in the docent-default condition; G1--G12 are anonymised participant IDs.}
\label{tab:app-g-demo}\\
\hline
\textbf{ID} & \textbf{Gender} & \textbf{Age} & \textbf{Nationality} & \textbf{Guide lang.} & \textbf{Occupation} & \textbf{XR exp.} \\
\hline
\endfirsthead

\hline
\textbf{ID} & \textbf{Gender} & \textbf{Age} & \textbf{Nationality} & \textbf{Guide lang.} & \textbf{Occupation} & \textbf{XR exp.} \\
\hline
\endhead

G1  & Male   & 19 & Hong Kong, China & Cantonese & UG student               & Some \\
\hline
G2  & Female & 19 & Hong Kong, China & English   & UG student               & Almost none \\
\hline
G3  & Male   & 23 & Chinese          & Mandarin  & Technology entrepreneur  & Very extensive \\
\hline
G4  & Female & 39 & USA              & English   & Creative practitioner    & Very extensive \\
\hline
G5  & Male   & 66 & Chinese          & Mandarin  & Retired visitor          & Almost none \\
\hline
G6  & Female & 50 & Chinese          & Mandarin  & Finance professional     & Some \\
\hline
G7  & Female & 19 & Belgian          & English   & Undergraduate student    & Some \\
\hline
G8  & Male   & 52 & Chinese          & Mandarin  & Teacher                  & Some \\
\hline
G9  & Male   & 55 & Chinese          & Mandarin  & Academic                 & Very extensive \\
\hline
G10 & Female & 24 & Chinese          & Mandarin  & Graduate student         & Very extensive \\
\hline
G11 & Female & 22 & Chinese          & Mandarin  & Creative practitioner    & Some \\
\hline
G12 & Male   & 20 & Chinese          & Cantonese & Undergraduate student    & Some \\
\hline

\end{longtable}
\endgroup

\section{Condition Briefing and Docent Preparation}
\label{app:condition-details}

Both conditions used the same in-headset tutorial for language selection, label access, Dream-Butterfly summoning, and microphone/listening state. A separate condition-specific briefing established the default source of artwork interpretation. In the agent-default condition, participants were told that Dream-Butterfly was the default source for artwork interpretation. They were also explicitly told that the accompanying human docent remained available for any question or need outside artwork interpretation, including route choices, safety or comfort concerns, and technical problems. During these sessions, docents avoided routine artwork interpretation except when needed to resolve a breakdown or keep the visit coherent. In the docent-default condition, participants were told that the human docent would provide the main narration and answer artwork questions; Dream-Butterfly remained available as an optional channel.

Four docents covered the study sessions. Three could guide in both Mandarin and English, and one was assigned to Cantonese-language sessions. Every docent completed two rounds of training. The training covered all participating artists and the curatorial material published on the exhibition labels. Docents were expected to give a basic account of each work's artist, creative motivation, and main idea, and their preparation was bounded to this label-level material. The training source pack was the same corpus stored in Dream-Butterfly's retrieval-augmented generation system.

Communication records confirmed that 9 of 12 docent-default participants invoked Dream-Butterfly at least once; the paper's account of their reasons draws on post-visit interviews.

\section{Post-study questionnaire}
\label{app:questionnaire}

\renewcommand{\arraystretch}{1.15}
\small
\setlength{\tabcolsep}{3pt}

\subsection{Custom 7-point Likert items used in both conditions}
\begin{longtable}{|p{0.10\linewidth}|p{0.22\linewidth}|p{0.63\linewidth}|}
\caption{Custom items administered in both conditions. Response scale: 1 (Strongly disagree) -- 7 (Strongly agree). Items marked with \emph{(reversed)} were reverse-scored.}
\label{tab:app-custom}\\
\hline
\textbf{\#} & \textbf{Construct} & \textbf{Item wording}\\
\hline
\endfirsthead

\hline
\textbf{\#} & \textbf{Construct} & \textbf{Item wording}\\
\hline
\endhead

1  & Explanation Access \& Interaction & I could access explanations easily whenever I wanted. \\
\hline
2  & Explanation Access \& Interaction & I could naturally ask follow-up questions / clarifications and receive responses. \\
\hline
3  & Explanation Access \& Interaction & The explanations were closely tied to what I was seeing in-situ (context-relevant). \\
\hline
4  & Explanation Access \& Interaction & Getting explanations interrupted my viewing or immersion. \emph{(reversed)} \\
\hline

5  & Immersion \& Engagement & I stayed focused for most of the experience. \\
\hline
6  & Immersion \& Engagement & I felt absorbed / drawn into the experience. \\
\hline
7  & Immersion \& Engagement & I was often distracted by external factors (crowds, walking, device operation, weather, etc.). \emph{(reversed)} \\
\hline
8  & Immersion \& Engagement & Overall, this experience made me more engaged with the artworks. \\
\hline

9  & TLX-mini (workload) & I had to put in a lot of effort to complete the visit and access explanations. \\
\hline
10 & TLX-mini (workload) & I felt frustrated/irritated while accessing explanations or following the guiding process. \\
\hline

11 & Role Clarity \& Handoff & I clearly understood what the primary guide (staff or Dream-Butterfly) could help me with. \\
\hline
12 & Role Clarity \& Handoff & When I had an issue, it was easy to decide who to turn to first. \\
\hline
13 & Role Clarity \& Handoff & Switching between staff, Dream-Butterfly, and myself was smooth. \\
\hline
14 & Role Clarity \& Handoff & I felt confused or lost time because responsibilities were unclear. \emph{(reversed)} \\
\hline
15 & Role Clarity \& Handoff & I ended up carrying more guiding responsibility than I expected or wanted. \emph{(reversed)} \\
\hline

\end{longtable}

\subsection{UEQ-S (semantic differential; both conditions)}
\begin{longtable}{|p{0.10\linewidth}|p{0.36\linewidth}|p{0.36\linewidth}|p{0.12\linewidth}|}
\caption{UEQ-S items (semantic differential). Response format: 1 = closer to the left adjective; 7 = closer to the right adjective \cite{schrepp2017ueqs}.}
\label{tab:app-ueqs}\\
\hline
\textbf{\#} & \textbf{Left adjective} & \textbf{Right adjective} & \textbf{Scale}\\
\hline
\endfirsthead

\hline
\textbf{\#} & \textbf{Left adjective} & \textbf{Right adjective} & \textbf{Scale}\\
\hline
\endhead

16 & Obstructive & Supportive & 1--7\\
\hline
17 & Complicated & Easy & 1--7\\
\hline
18 & Inefficient & Efficient & 1--7\\
\hline
19 & Confusing & Clear & 1--7\\
\hline
20 & Boring & Exciting & 1--7\\
\hline
21 & Not interesting & Interesting & 1--7\\
\hline
22 & Conventional & Inventive & 1--7\\
\hline
23 & Usual & Leading edge & 1--7\\
\hline

\end{longtable}

\subsection{Responsibility Distribution Matrix (both conditions)}
\begin{longtable}{|p{0.10\linewidth}|p{0.70\linewidth}|p{0.16\linewidth}|}
\caption{Responsibility Distribution Matrix. For each domain, participants allocated 100 points across three parties: Human staff, Dream-Butterfly, and \textit{Virtual labels / Myself \& companions}. The three numbers must sum to 100.}
\label{tab:app-rdm}\\
\hline
\textbf{\#} & \textbf{Guiding responsibility domain} & \textbf{Format}\\
\hline
\endfirsthead

\hline
\textbf{\#} & \textbf{Guiding responsibility domain} & \textbf{Format}\\
\hline
\endhead

24 & Navigation \& wayfinding (finding artworks and reaching the next point). & 0--100 allocation\\
\hline
25 & Explaining artwork content \& answering questions. & 0--100 allocation\\
\hline
26 & Pacing (time spent, depth of explanation, skipping). & 0--100 allocation\\
\hline
27 & Troubleshooting MR/device issues (e.g., setup, failures, localization). & 0--100 allocation\\
\hline
28 & Safety \& comfort (positioning, obstacles, stamina, weather). & 0--100 allocation\\
\hline

\end{longtable}

\section{Additional quantitative reporting}
\label{app:quant-reporting}

\subsection{Complete Mann--Whitney U results}
All comparisons use two-sided asymptotic Mann--Whitney U tests with tie and continuity correction. $U$ is reported for the agent-default group, and the absolute effect size is $r=|z|/\sqrt{N}$, where $N=24$. Each condition contains 12 participants.

\setlength{\tabcolsep}{4pt}
\footnotesize
\begin{longtable}{p{0.28\linewidth}>{\centering\arraybackslash}p{0.16\linewidth}>{\centering\arraybackslash}p{0.16\linewidth}rrr}
\caption{Complete between-condition results. Mdn ($n$) gives the median and valid sample size for the agent-default and docent-default conditions.}
\label{tab:app-mwu}\\
\toprule
\textbf{Outcome} & \textbf{Agent-default Mdn ($n$)} & \textbf{Docent-default Mdn ($n$)} & \textbf{$U_{\mathrm{agent\mbox{-}default}}$} & \textbf{$p$} & \textbf{$r$} \\
\midrule
\endfirsthead
\toprule
\textbf{Outcome} & \textbf{Agent-default Mdn ($n$)} & \textbf{Docent-default Mdn ($n$)} & \textbf{$U_{\mathrm{agent\mbox{-}default}}$} & \textbf{$p$} & \textbf{$r$} \\
\midrule
\endhead
Explanation Access \& Interaction & 5.63 (12) & 3.75 (12) & 128.0 & .00127 & .66 \\
Immersion \& Engagement & 5.50 (12) & 4.38 (12) & 124.0 & .00276 & .61 \\
UEQ-S Pragmatic & 1.25 (12) & 0.88 (12) & 91.0 & .28295 & .22 \\
UEQ-S Hedonic & 2.13 (12) & $-0.88$ (12) & 142.5 & .000049 & .83 \\
TLX-mini Effort & 2.00 (12) & 1.50 (12) & 75.5 & .85506 & .04 \\
TLX-mini Frustration & 2.00 (12) & 1.50 (12) & 78.0 & .73059 & .07 \\
Role capability clarity & 6.00 (12) & 5.00 (12) & 133.5 & .000223 & .75 \\
First-resort clarity & 6.00 (12) & 4.00 (12) & 137.5 & .000110 & .79 \\
Switching smoothness & 6.00 (12) & 4.00 (12) & 116.0 & .00955 & .53 \\
Less role confusion & 4.00 (12) & 3.50 (12) & 79.0 & .70196 & .08 \\
Less unexpected burden & 2.00 (12) & 4.00 (12) & 34.0 & .02807 & .45 \\
Role Clarity \& Handoff composite & 4.90 (12) & 4.00 (12) & 119.0 & .00691 & .55 \\
\bottomrule
\end{longtable}

\newpage
\small
\subsection{Responsibility-allocation variability}
Table~\ref{tab:app-rdm-variability} reports the means visualised in the 100\% stacked chart together with between-participant SDs. Ten of the 120 participant-by-domain allocations summed to 99 because of integer rounding; for the chart and summary statistics, those three-party rows were proportionally normalised to 100. The source responses were retained unchanged.

\setlength{\LTpre}{2pt}
\setlength{\LTpost}{2pt}
\captionsetup[table]{skip=4pt}
\renewcommand{\arraystretch}{1.0}
\footnotesize
\begin{longtable}{p{0.18\linewidth}p{0.20\linewidth}rrr}
\caption{Responsibility allocations as mean percentage (SD), $n=12$ per condition.}
\label{tab:app-rdm-variability}\\
\toprule
\textbf{Domain} & \textbf{Condition} & \textbf{Human staff} & \textbf{Dream-Butterfly} & \textbf{Labels/Self} \\
\midrule
\endfirsthead
\toprule
\textbf{Domain} & \textbf{Condition} & \textbf{Human staff} & \textbf{Dream-Butterfly} & \textbf{Labels/Self} \\
\midrule
\endhead
Navigation & Agent-default & 69.2 (28.1) & 26.7 (25.6) & 4.2 (6.3) \\
Navigation & Docent-default & 87.5 (13.6) & 5.4 (7.8) & 7.1 (11.4) \\
Explaining & Agent-default & 2.7 (3.9) & 90.8 (7.5) & 6.5 (5.3) \\
Explaining & Docent-default & 54.2 (21.5) & 24.2 (17.3) & 21.7 (12.7) \\
Pacing & Agent-default & 48.9 (18.8) & 24.3 (13.1) & 26.8 (13.7) \\
Pacing & Docent-default & 72.3 (23.3) & 11.8 (14.7) & 15.9 (15.0) \\
Troubleshooting & Agent-default & 75.3 (27.7) & 6.1 (9.8) & 18.6 (23.5) \\
Troubleshooting & Docent-default & 93.3 (15.0) & 4.2 (14.4) & 2.5 (6.2) \\
Safety & Agent-default & 60.8 (23.0) & 16.7 (16.0) & 22.5 (17.5) \\
Safety & Docent-default & 82.5 (21.8) & 6.7 (16.1) & 10.8 (16.2) \\
\bottomrule
\end{longtable}

\small
\subsection{Guiding-language distribution}
Guiding language was selected by participants rather than assigned. Counts were Mandarin: 13 (agent-default: 6; docent-default: 7), English: 7 (agent-default: 4; docent-default: 3), and Cantonese: 4 (agent-default: 2; docent-default: 2). Because condition-by-language cells contain only 2--7 participants, language-stratified outcome summaries are exploratory and are not used for inferential or causal claims.

\normalsize
\setlength{\tabcolsep}{6pt}

\subsection{SUS (agent-default condition only)}
\begin{longtable}{|p{0.10\linewidth}|p{0.84\linewidth}|}
\caption{System Usability Scale (SUS) items for the Dream-Butterfly guide (agent-default condition only). Response scale: 1 (Strongly disagree) -- 5 (Strongly agree). Items marked with \emph{(reversed)} were reverse-scored \cite{brooke1996sus}.}
\label{tab:app-sus}\\
\hline
\textbf{\#} & \textbf{Item wording}\\
\hline
\endfirsthead

\hline
\textbf{\#} & \textbf{Item wording}\\
\hline
\endhead

29 & I think that I would like to use this system frequently. \\
\hline
30 & I found the system unnecessarily complex. \emph{(reversed)} \\
\hline
31 & I thought the system was easy to use. \\
\hline
32 & I think that I would need the support of a technical person to be able to use this system. \emph{(reversed)} \\
\hline
33 & I found the various functions in this system were well integrated. \\
\hline
34 & I thought there was too much inconsistency in this system. \emph{(reversed)} \\
\hline
35 & I would imagine that most people would learn to use this system very quickly. \\
\hline
36 & I found the system very cumbersome to use. \emph{(reversed)} \\
\hline
37 & I felt very confident using the system. \\
\hline
38 & I needed to learn a lot of things before I could get going with this system. \emph{(reversed)} \\
\hline

\end{longtable}

% -----------------------------
% Appendix C: Motion Equations
% -----------------------------
\section{Motion control equations for the embodied following trajectory}
\label{app:motion-eq}

For reproducibility, we provide the formal definitions used by the world-space following behaviour and wing-flap speed modulation described in the main paper's system section.

\begin{equation}
\label{eq:goalpoint}
p^*(t)=
\begin{cases}
T_{\text{cam}}(t)\,o_{\text{follow}} + (0,\,A\sin(\omega t),\,0) & \text{(free flight / grip not held)}\\
T_{\text{hand}}(t)\,o_{\text{land}} & \text{(grip held / landing)}
\end{cases}
\end{equation}

Here, $o_{\text{follow}}$ and $o_{\text{land}}$ are designer-authored offsets editable in the inspector.
When the grip is not held, the sinusoidal term adds a subtle hovering cue to preserve discoverability without making the agent feel rigidly locked to the view.

\begin{equation}
\label{eq:flapspeed}
\alpha(t)=
\begin{cases}
\alpha_{\text{land}} & \text{if grip held and }\|p^*(t) - p(t)\|\le\varepsilon\\
\alpha_{\text{acc}} & \text{if }t-t_{\text{switch}}<0.5\text{ s}\\
\alpha_{\text{cruise}} & \text{otherwise}
\end{cases}
\end{equation}

The brief $\alpha_{\text{acc}}$ window makes state changes perceptible as a short ``start-up'' burst; the cruise rate sustains readability during pursuit; and the slower landing rate conveys rest and readiness for dialogue.

%% file: ref.bib
@inproceedings{congdon2023mixedagents,
  author    = {Congdon, Ben J. and Park, Gun Woo (Warren) and Zhang, Jingyi and Steed, Anthony},
  title     = {Comparing Mixed Reality Agent Representations: Studies in the Lab and in the Wild},
  booktitle = {29th ACM Symposium on Virtual Reality Software and Technology},
  series    = {VRST 2023},
  pages     = {1--11},
  publisher = {ACM},
  year      = {2023},
  doi       = {10.1145/3611659.3615719},
  url       = {https://doi.org/10.1145/3611659.3615719}
}

@inproceedings{wells2025llmxr,
  author    = {Wells, Brody and Wang, Nanjia and Maurer, Frank},
  title     = {When to Call the {LLM}: Adaptive Speech Interfaces for Natural Language Input in eXtended Reality Environments},
  booktitle = {Proceedings of the 2025 31st ACM Symposium on Virtual Reality Software and Technology},
  series    = {VRST '25},
  pages     = {80:1--80:10},
  publisher = {ACM},
  year      = {2025},
  doi       = {10.1145/3756884.3766015},
  url       = {https://doi.org/10.1145/3756884.3766015}
}

@inproceedings{schott2024guideapprentices,
  author    = {Schott, Ephraim and L{\'o}pez Garc{\'i}a, Irene and Zoeppig, Tony Jan and Hartmann, Manuel and Froehlich, Bernd},
  title     = {The Guide's Apprentices: Engaging Visitors of Virtual Museums through Appropriate Agent Embodiments},
  booktitle = {30th ACM Symposium on Virtual Reality Software and Technology},
  series    = {VRST '24},
  pages     = {1--2},
  publisher = {ACM},
  year      = {2024},
  doi       = {10.1145/3641825.3689529},
  url       = {https://doi.org/10.1145/3641825.3689529}
}

@inproceedings{cascarano2025seehear,
  author    = {Cascarano, Pasquale and Loretti, Andrea and Zanuttini, Luca and Giunchi, Daniele and Bovo, Riccardo and Hajahmadi, Shirin and Vallasciani, Giacomo and Martinoni, Matteo and Marfia, Gustavo},
  title     = {See It and Hear It: Multimodal Guidance in {MR}-Based Neurosurgical Simulation for Skill Retention},
  booktitle = {Proceedings of the 2025 31st ACM Symposium on Virtual Reality Software and Technology},
  series    = {VRST '25},
  pages     = {1--11},
  publisher = {ACM},
  year      = {2025},
  doi       = {10.1145/3756884.3766034},
  url       = {https://doi.org/10.1145/3756884.3766034}
}

@inproceedings{collins2026llmvrguide,
  author    = {Collins, Jazmin and Lin, Sharon Y. and Liu, Tianqi and Won, Andrea Stevenson and Azenkot, Shiri},
  title     = {Understanding the Use of a Large Language Model-Powered Guide to Make Virtual Reality Accessible for Blind and Low Vision People},
  booktitle = {Proceedings of the 2026 CHI Conference on Human Factors in Computing Systems},
  series    = {CHI '26},
  pages     = {1115:1--1115:16},
  publisher = {ACM},
  year      = {2026},
  doi       = {10.1145/3772318.3791143},
  url       = {https://doi.org/10.1145/3772318.3791143}
}

@inproceedings{liu2026agenthands,
  author    = {Liu, Ziyi and Li, David and Zhou, Zhongyi and Kim, David and Du, Ruofei and Qian, Xun},
  title     = {AgentHands: Generating Interactive Hand Gestures for Spatially Grounded Agent Conversations in {XR}},
  booktitle = {Proceedings of the 2026 CHI Conference on Human Factors in Computing Systems},
  series    = {CHI '26},
  pages     = {1520:1--1520:24},
  publisher = {ACM},
  year      = {2026},
  doi       = {10.1145/3772318.3790938},
  url       = {https://doi.org/10.1145/3772318.3790938}
}

@inproceedings{cheng2026auditoryagents,
  author    = {Cheng, Yi Fei and Bloch, Jarod and Wang, Alexander and Bianchi, Andrea and Withana, Anusha and Guo, Anhong and Heller, Laurie M. and Lindlbauer, David},
  title     = {Auditorily Embodied Conversational Agents: Effects of Spatialization and Situated Audio Cues on Presence and Social Perception},
  booktitle = {Proceedings of the 2026 CHI Conference on Human Factors in Computing Systems},
  series    = {CHI '26},
  pages     = {1465:1--1465:16},
  publisher = {ACM},
  year      = {2026},
  doi       = {10.1145/3772318.3791794},
  url       = {https://doi.org/10.1145/3772318.3791794}
}

@inproceedings{wang2026aispeakfirst,
  author    = {Wang, Duo and Song, Wenxuan and Ni, Jianwei and Zheng, Qingxiao and Zhou, Yuhan and Liang, Kyrian and Yao, Mike and Cao, Caroline G. L.},
  title     = {Should the {AI} Speak First? Evaluating Proactive vs. Reactive Facilitation in Mixed-Reality Medical Training},
  booktitle = {Proceedings of the 2026 CHI Conference on Human Factors in Computing Systems},
  series    = {CHI '26},
  pages     = {1631:1--1631:19},
  publisher = {ACM},
  year      = {2026},
  doi       = {10.1145/3772318.3791910},
  url       = {https://doi.org/10.1145/3772318.3791910}
}

@inproceedings{madritsch2026arconversation,
  author    = {Madritsch, Juliana H. and Duricic, Tomislav and ElSayed, Neven A. M. and Kopeinik, Simone and Veas, Eduardo E.},
  title     = {{AI}-Powered Conversational Assistance in Augmented Reality for Multi-Step Tasks},
  booktitle = {2026 IEEE Conference on Virtual Reality and 3D User Interfaces (VR)},
  pages     = {519--529},
  publisher = {IEEE},
  year      = {2026},
  doi       = {10.1109/VR67842.2026.00072},
  url       = {https://doi.org/10.1109/VR67842.2026.00072}
}

@inproceedings{kaschub2026referentialassistant,
  author    = {Kaschub, Lina and V{\"o}lckers, Bado and Turhan, Ugur and Huesmann, Philipp and Kruse, Lucie and Steinicke, Frank},
  title     = {Comparing Referential Interactions with an Intelligent Assistant in Virtual Reality},
  booktitle = {2026 IEEE Conference on Virtual Reality and 3D User Interfaces (VR)},
  pages     = {453--463},
  publisher = {IEEE},
  year      = {2026},
  doi       = {10.1109/VR67842.2026.00066},
  url       = {https://doi.org/10.1109/VR67842.2026.00066}
}

@article{zhu2026genaixr,
  author = {Zhu, Mingyu and Chen, Jiangong and Li, Bin},
  title = {When Generative AI Meets Extended Reality: Enabling Scalable and Natural Interactions},
  journal = {IEEE Internet Computing},
  year = {2026},
  doi = {10.1109/MIC.2025.3619462}
}

@misc{yu2026realitycopilot,
  author = {Yu, Liuchuan and Zhang, Yongqi and Yu, Lap-Fai},
  title = {Reality Copilot: Voice-First Human-AI Collaboration in Mixed Reality Using Large Multimodal Models},
  year = {2026},
  eprint = {2602.11025},
  archivePrefix = {arXiv},
  primaryClass = {cs.HC},
  doi = {10.48550/arXiv.2602.11025},
  url = {https://arxiv.org/abs/2602.11025}
}

@misc{sun2026jarvis,
  author = {Sun, Yusi and Jiang, Ying and Lu, Jiayin and Yang, Yin and Kuo, Yong-Hong and Jiang, Chenfanfu},
  title = {JARVIS: A Just-in-Time AR Visual Instruction System for Cross-Reality Task Guidance},
  year = {2026},
  eprint = {2604.10108},
  archivePrefix = {arXiv},
  primaryClass = {cs.HC},
  doi = {10.48550/arXiv.2604.10108},
  url = {https://arxiv.org/abs/2604.10108}
}

@misc{ihara2026cinemaworld,
  author = {Ihara, Keiichi and Lee, DaeHo and Abe, Manato and Jo, Hye-Young and Suzuki, Ryo},
  title = {CinemaWorld: Generative Augmented Reality with LLMs and 3D Scene Generation for Movie Augmentation},
  year = {2026},
  eprint = {2603.08060},
  archivePrefix = {arXiv},
  primaryClass = {cs.HC},
  doi = {10.48550/arXiv.2603.08060},
  url = {https://arxiv.org/abs/2603.08060}
}

@article{morse2021casual,
  title={Casual leisure in rich-prospect: Advancing visual information behavior for digital museum collections},
  author={Morse, Christopher and Niess, Jasmin and Lallemand, Carine and Wieneke, Lars and Koenig, Vincent},
  journal={Journal on Computing and Cultural Heritage (JOCCH)},
  volume={14},
  number={3},
  pages={1--23},
  year={2021},
  publisher={ACM New York, NY, USA}
}

@inproceedings{roberts2018digital,
  title={Digital exhibit labels in museums: promoting visitor engagement with cultural artifacts},
  author={Roberts, Jessica and Banerjee, Amartya and Hong, Annette and McGee, Steven and Horn, Michael and Matcuk, Matt},
  booktitle={Proceedings of the 2018 CHI Conference on Human Factors in Computing Systems},
  pages={1--12},
  year={2018}
}

@inproceedings{kopp2005museumguide,
  title     = {A Conversational Agent as Museum Guide--Design and Evaluation of a Real-World Application},
  author    = {Kopp, Stefan and Gesellensetter, Lars and Kr{\"a}mer, Nicole C. and Wachsmuth, Ipke},
  booktitle = {Intelligent Virtual Agents (IVA 2005)},
  series    = {Lecture Notes in Computer Science},
  publisher = {Springer},
  doi       = {10.1007/11550617_28},
  year      = {2005}
}

@inproceedings{aoki2002sottovoce,
  author    = {Paul M. Aoki and
               Rebecca E. Grinter and
               Amy Hurst and
               Margaret H. Szymanski and
               James D. Thornton and
               Allison Woodruff},
  title     = {{Sotto Voce}: Exploring the Interplay of Conversation and Mobile Audio Spaces},
  booktitle = {Proceedings of the SIGCHI Conference on Human Factors in Computing Systems (CHI '02)},
  year      = {2002},
  pages     = {431--438},
  publisher = {ACM},
  doi       = {10.1145/503376.503454}
}

@article{cameron1971templeforum,
  author  = {Cameron, Duncan F.},
  title   = {The Museum, a Temple or the Forum},
  journal = {Curator: The Museum Journal},
  volume  = {14},
  number  = {1},
  pages   = {11--24},
  year    = {1971},
  doi     = {10.1111/j.2151-6952.1971.tb00416.x}
}

@book{goffman1981formsoftalk,
  author    = {Goffman, Erving},
  title     = {Forms of Talk},
  publisher = {University of Pennsylvania Press},
  address   = {Philadelphia, PA},
  year      = {1981},
  isbn      = {9780812211122}
}

@inproceedings{su2025simviews,
  author    = {Su, Mingyang and Liu, Chao and Zhang, Jingling and Wu, Shuang and Fan, Mingming},
  title     = {SimViews: An Interactive Multi-Agent System Simulating Visitor-to-Visitor Conversational Patterns to Present Diverse Perspectives of Artifacts in Virtual Museums},
  booktitle = {Proceedings of the 33rd ACM International Conference on Multimedia (MM '25)},
  year      = {2025},
  pages     = {6740--6749},
  publisher = {Association for Computing Machinery},
  doi       = {10.1145/3746027.3754593}
}

@inproceedings{grinter2002revisiting,
  author    = {Rebecca E. Grinter and
               Paul M. Aoki and
               Amy Hurst and
               Margaret H. Szymanski and
               James D. Thornton and
               Allison Woodruff},
  title     = {Revisiting the Visit: Understanding How Technology Can Shape the Museum Visit},
  booktitle = {Proceedings of the 2002 ACM Conference on Computer Supported Cooperative Work (CSCW '02)},
  year      = {2002},
  pages     = {146--155},
  publisher = {ACM},
  doi       = {10.1145/587078.587100}
}

@inproceedings{woodruff2001guidebooks,
  author    = {Allison Woodruff and
               Paul M. Aoki and
               Amy Hurst and
               Margaret H. Szymanski},
  title     = {The Conversational Role of Electronic Guidebooks},
  booktitle = {Ubiquitous Computing (UbiComp 2001)},
  series    = {Lecture Notes in Computer Science},
  volume    = {2201},
  year      = {2001},
  pages     = {187--208},
  publisher = {Springer}
}

@article{lewis2020rag,
  author  = {Patrick Lewis and Ethan Perez and Aleksandra Piktus and Fabio Petroni and
             Vladimir Karpukhin and Naman Goyal and Heinrich K{\"u}ttler and Mike Lewis and
             Wen{-}tau Yih and Tim Rockt{\"a}schel and Sebastian Riedel and Douwe Kiela},
  title   = {Retrieval-Augmented Generation for Knowledge-Intensive NLP Tasks},
  journal = {Advances in Neural Information Processing Systems (NeurIPS)},
  year    = {2020},
  url     = {https://arxiv.org/abs/2005.11401}
}

@article{trichopoulos2023chatgpt4guide,
  title   = {Crafting a Museum Guide Using {ChatGPT}-4},
  author  = {Trichopoulos, Georgios and Konstantakis, Markos and Caridakis, George and Katifori, Akrivi and Koukouli, Myrto},
  journal = {Big Data and Cognitive Computing},
  volume  = {7},
  number  = {3},
  pages   = {148},
  year    = {2023},
  doi     = {10.3390/bdcc7030148}
}

@article{braun2019reflecting,
  title={Reflecting on reflexive thematic analysis},
  author={Braun, Virginia and Clarke, Victoria},
  journal={Qualitative research in sport, exercise and health},
  volume={11},
  number={4},
  pages={589--597},
  year={2019},
  publisher={Taylor \& Francis}
}

@incollection{brooke1996sus,
  author    = {Brooke, John},
  title     = {SUS: A ``Quick and Dirty'' Usability Scale},
  booktitle = {Usability Evaluation in Industry},
  editor    = {Jordan, Patrick W. and Thomas, Bruce and Weerdmeester, Bernard A. and McClelland, Ian L.},
  pages     = {189--194},
  publisher = {Taylor \& Francis},
  year      = {1996}
}

@article{schrepp2017ueqs,
  author  = {Schrepp, Martin and Hinderks, Andreas and Thomaschewski, J{\"o}rg},
  title   = {Design and Evaluation of a Short Version of the User Experience Questionnaire (UEQ-S)},
  journal = {International Journal of Interactive Multimedia and Artificial Intelligence},
  volume  = {4},
  number  = {6},
  pages   = {103--108},
  year    = {2017}
}

@incollection{hart1988nasatlx,
  author    = {Hart, Sandra G. and Staveland, Lowell E.},
  title     = {Development of NASA-TLX (Task Load Index): Results of Empirical and Theoretical Research},
  booktitle = {Human Mental Workload},
  editor    = {Hancock, Peter A. and Meshkati, Najmedin},
  series    = {Advances in Psychology},
  volume    = {52},
  pages     = {139--183},
  publisher = {North-Holland},
  year      = {1988}
}

@article{best2012making,
  title={Making museum tours better: Understanding what a guided tour really is and what a tour guide really does},
  author={Best, Katie},
  journal={Museum Management and Curatorship},
  volume={27},
  number={1},
  pages={35--52},
  year={2012},
  publisher={Taylor \& Francis}
}

@inproceedings{10.1145/3769534.3769544,
author = {Bauer, Axel and Hagler, Juergen and Kocur, Martin},
title = {Interactive Augmented Reality Experiences for Urban Art Spaces Using Markerless Tracking},
year = {2025},
isbn = {9798400718458},
publisher = {Association for Computing Machinery},
address = {New York, NY, USA},
url = {https://doi.org/10.1145/3769534.3769544},
doi = {10.1145/3769534.3769544},
booktitle = {Proceedings of the 18th International Symposium on Visual Information Communication and Interaction},
articleno = {46},
numpages = {6},
location = {
},
series = {VINCI '25}
}

@article{harrison2025making,
  title={Making Meaningful Places: Augmented Reality for Digital Placemaking in Clearwater, Florida},
  author={Harrison, Laura K and Kaplan, Howard and Matthews, Jennifer T and Santiago, Eric},
  journal={ACM Journal on Computing and Cultural Heritage},
  volume={18},
  number={1},
  pages={1--21},
  year={2025},
  publisher={ACM New York, NY},
  doi={10.1145/3703916},
  url={https://doi.org/10.1145/3703916}
}

@article{knabe2025lucky,
  title={Lucky Coincidences: Experiencing Serendipity in Museums and Beyond},
  author={Knabe, Max and Altenm{\"u}ller, Marlene Sophie and Kampschulte, Lorenz and Bergunde, Franca and Hahn, Nathalie and J{\"u}ngert, Anne and Kammermeier, Katja and McTassney, Michelle and Ruhdorfer, Benedikt and Weber, Emily},
  journal={Journal of Applied Social Psychology},
  volume={55},
  number={11},
  pages={855--870},
  year={2025},
  publisher={Wiley Online Library},
  doi={10.1111/jasp.70020},
  url={https://doi.org/10.1111/jasp.70020}
}

@inproceedings{10972673,
  author = {Bauer, Axel and Kocur, Martin and Hagler, Juergen},
  title = {Augmenting Murals: Creating Playful AR Experiences for Wall Art Exhibitions},
  booktitle = {2025 IEEE Conference on Virtual Reality and 3D User Interfaces Abstracts and Workshops (VRW)},
  year = {2025},
  pages = {8--12},
  publisher = {IEEE},
  address = {Piscataway, NJ, USA},
  doi = {10.1109/VRW66409.2025.00009},
  url = {https://doi.org/10.1109/VRW66409.2025.00009}
}

@article{macario2022comprehensive,
  title={A comprehensive survey of visual slam algorithms},
  author={Macario Barros, Andr{\'e}a and Michel, Maugan and Moline, Yoann and Corre, Gwenol{\'e} and Carrel, Fr{\'e}d{\'e}rick},
  journal={Robotics},
  volume={11},
  number={1},
  pages={24},
  year={2022},
  publisher={MDPI}
}

@article{hein1999meaning,
  title={Is meaning making constructivism? Is constructivism meaning making},
  author={Hein, George},
  journal={The Exhibitionist},
  volume={18},
  number={2},
  pages={15--18},
  year={1999}
}

@article{matviienko2022arsightseeing,
  author  = {Andrii Matviienko and Sebastian G{\"{u}}nther and Sebastian Ritzenhofen and Max M{\"{u}}hlh{\"{a}}user},
  title   = {{AR} Sightseeing: Comparing Information Placements at Outdoor Historical Heritage Sites using Augmented Reality},
  journal = {Proc. {ACM} Hum. Comput. Interact.},
  volume  = {6},
  number  = {{MHCI}},
  pages   = {1--17},
  year    = {2022},
  doi     = {10.1145/3546729},
  url     = {https://doi.org/10.1145/3546729}
}

@inproceedings{ghaemi2023placement,
  author    = {Zeinab Ghaemi and Kadek Ananta Satriadi and Ulrich Engelke and Barrett Ens and Bernhard Jenny},
  editor    = {Tony Huang and Misha Sra and Ferran Argelaguet and Pedro Lopes and Mayra Donaji Barrera Machuca},
  title     = {Visualization Placement for Outdoor Augmented Data Tours},
  booktitle = {Proceedings of the 2023 {ACM} Symposium on Spatial User Interaction,
               {SUI} 2023, Sydney, NSW, Australia, October 13-15, 2023},
  pages     = {9:1--9:14},
  publisher = {{ACM}},
  year      = {2023},
  doi       = {10.1145/3607822.3614518},
  url       = {https://doi.org/10.1145/3607822.3614518}
}

@inproceedings{wang2019agents,
  author    = {Isaac Wang and Jesse Smith and Jaime Ruiz},
  editor    = {Stephen A. Brewster and Geraldine Fitzpatrick and Anna L. Cox and Vassilis Kostakos},
  title     = {Exploring Virtual Agents for Augmented Reality},
  booktitle = {Proceedings of the 2019 {CHI} Conference on Human Factors in Computing Systems,
               {CHI} 2019, Glasgow, Scotland, UK, May 04-09, 2019},
  pages     = {281},
  publisher = {{ACM}},
  year      = {2019},
  doi       = {10.1145/3290605.3300511},
  url       = {https://doi.org/10.1145/3290605.3300511}
}

@inproceedings{huang2022proxemics,
  author    = {Ann Huang and Pascal Knierim and Francesco Chiossi and Lewis L. Chuang and Robin Welsch},
  title     = {Proxemics for Human-Agent Interaction in Augmented Reality},
  booktitle = {Proceedings of the 2022 {CHI} Conference on Human Factors in Computing Systems ({CHI} 2022)},
  pages     = {421:1--421:13},
  publisher = {ACM},
  year      = {2022},
  doi       = {10.1145/3491102.3517593},
  url       = {https://doi.org/10.1145/3491102.3517593}
}

@inproceedings{Lai2022ConditionalDelegation,
  author    = {Lai, Vivian and Liu, Han and Liao, Q. Vera and Yuan, Yiwei and Tan, Chenhao},
  title     = {Human-AI Collaboration via Conditional Delegation: A Case Study of Content Moderation},
  booktitle = {Proceedings of the 2022 CHI Conference on Human Factors in Computing Systems (CHI '22)},
  year      = {2022},
  publisher = {Association for Computing Machinery},
  address   = {New York, NY, USA},
  doi       = {10.1145/3491102.3501999}
}

@article{Moon2024CrossDeviceMRArt,
  title   = {Mixed-reality art as shared experience for cross-device users: Materialize, understand, and explore},
  author  = {Moon, Hayoun and Saade, Mia and Enriquez, Daniel and Duer, Zachary and Moon, Hye Sung and Lee, Sang Won and Jeon, Myounghoon},
  journal = {International Journal of Human-Computer Studies},
  volume  = {190},
  pages   = {103291},
  year    = {2024},
  month   = {10},
  doi     = {10.1016/j.ijhcs.2024.103291}
}

@article{Djuric2025PublicSpaceArtMRAR,
  title   = {Enhancing Public Space Experiences: Evaluating Mixed Reality and Augmented Reality as a Tool for Presenting Artworks in Public Spaces},
  author  = {{\DJ}uri{\'c}, V. and Kljaki{\'c}, M. and Krsti{\'c}, S.},
  journal = {Applied Sciences},
  volume  = {15},
  number  = {2},
  pages   = {870},
  year    = {2025},
  doi     = {10.3390/app15020870}
}

@article{li2025wenetspeech,
  title = {WenetSpeech-Yue: A Large-Scale Cantonese Speech Corpus with Multi-Dimensional Annotation},
  author = {Li, Longhao and Guo, Zhao and Chen, Hongjie and Dai, Yuhang and Zhang, Ziyu and Xue, Hongfei and Zuo, Tianlun and Wang, Chengyou and Wang, Shuiyuan and Li, Jie and others},
  journal = {Proceedings of the AAAI Conference on Artificial Intelligence},
  volume = {40},
  number = {37},
  pages = {31627--31635},
  year = {2026},
  doi = {10.1609/aaai.v40i37.40429},
  url = {https://doi.org/10.1609/aaai.v40i37.40429}
}

@inproceedings{10.1145/3719160.3736629,
author = {Reimann, Merle M. and Kunneman, Florian A. and Oertel, Catharine and Hindriks, Koen V.},
title = {Transparent Conversational Agents: The Impact of Capability Communication on User Behavior and Mental Model Alignment},
year = {2025},
isbn = {9798400715273},
publisher = {Association for Computing Machinery},
address = {New York, NY, USA},
url = {https://doi.org/10.1145/3719160.3736629},
doi = {10.1145/3719160.3736629},
booktitle = {Proceedings of the 7th ACM Conference on Conversational User Interfaces},
articleno = {48},
numpages = {12},
location = {
},
series = {CUI '25}
}

@article{DiChristofano2022AccentASRDisparities,
  title   = {Global Performance Disparities Between English-Language Accents in Automatic Speech Recognition},
  author  = {DiChristofano, Alex and Shuster, Henry and Chandra, Shefali and Patwari, Neal},
  journal = {arXiv preprint arXiv:2208.01157},
  year    = {2022},
  doi     = {10.48550/arXiv.2208.01157}
}

@article{Graham2024WhisperAccents,
  title   = {Evaluating OpenAI’s Whisper Automated Speech Recognition: analysis across diverse accents},
  author  = {Graham, C. and Roll, N.},
  journal = {JASA Express Letters},
  volume  = {4},
  number  = {2},
  pages   = {025206},
  year    = {2024},
  doi     = {10.1121/10.0024876}
}

@article{Wang2026WenetSpeechWu,
  title   = {WenetSpeech-Wu: Datasets, Benchmarks, and Models for a Unified Chinese Wu Dialect Speech Processing Ecosystem},
  author  = {Wang, Chengyou and Shao, Mingchen and Hu, Jingbin and Zhu, Zeyu and Xue, Hongfei and Mu, Bingshen and Xu, Xin and Duan, Xingyi and Zhang, Binbin and Zhu, Pengcheng and Ding, Chuang and Zhang, Xiaojun and Bu, Hui and Xie, Lei},
  journal = {arXiv preprint arXiv:2601.11027},
  year    = {2026},
  doi     = {10.48550/arXiv.2601.11027}
}

@inproceedings{Michel2025NotRepresentation,
  title={“It’s not a representation of me”: Examining Accent Bias and Digital Exclusion in Synthetic AI Voice Services},
  author={Michel, Shira and Kaur, Sufi and Gillespie, Sarah Elizabeth and Gleason, Jeffrey and Wilson, Christo and Ghosh, Avijit},
  booktitle={Proceedings of the 2025 ACM Conference on Fairness, Accountability, and Transparency},
  pages={228--245},
  year={2025}
}

@inproceedings{chen2025museumguideprefs,
author = {Chen, Bingqing and Wen, Ruoyu and Tan, Shufang and Li, Yue},
title = {Exploring User Preferences for Museum Guides: The Role of Chatbots in Shaping Interactive Experiences},
year = {2025},
isbn = {9798400713958},
publisher = {Association for Computing Machinery},
address = {New York, NY, USA},
url = {https://doi.org/10.1145/3706599.3720067},
doi = {10.1145/3706599.3720067},
booktitle = {Proceedings of the Extended Abstracts of the CHI Conference on Human Factors in Computing Systems},
articleno = {256},
numpages = {8},
location = {
},
series = {CHI EA '25}
}

@article{Koenecke2020RacialDisparitiesASR,
  title   = {Racial disparities in automated speech recognition},
  author  = {Koenecke, Allison and Nam, Andrew and Lake, Emily and Nudell, Joe and Quartey, Mingxuan and Mengesha, Zion and Toups, Connor and Rickford, John R. and Jurafsky, Dan and Goel, Sharad},
  journal = {Proceedings of the National Academy of Sciences},
  volume  = {117},
  number  = {14},
  pages   = {7684--7689},
  year    = {2020},
  doi     = {10.1073/pnas.1915768117}
}

@article{Wang2025DialectOlderAdultsAlignment,
  title   = {The effect of regional dialect variation on older adults’ interaction with computers: Syntactic alignment in a structured task and open-ended conversation},
  author  = {Wang, Yiran and Zhao, Jie and Zhao, Shengdong},
  journal = {International Journal of Human-Computer Studies},
  year    = {2025},
  doi     = {10.1016/j.ijhcs.2025.103473}
}

@article{Trefzger:2025:ymt,
author = {Stefanidi, Helen and S\"{u}nderkamp, Jan-Hendrik and Tatzgern, Markus and Itzlinger, Alina and Meschtscherjakov, Alexander},
title = {You're making things AR-kward: Exploring Augmented Reality In-the-Wild},
year = {2025},
issue_date = {September 2025},
publisher = {Association for Computing Machinery},
address = {New York, NY, USA},
volume = {9},
number = {5},
url = {https://doi.org/10.1145/3743740},
doi = {10.1145/3743740},
journal = {Proc. ACM Hum.-Comput. Interact.},
month = sep,
articleno = {MHCI041},
numpages = {24}
}

@article{sun2024hunyuan,
  title={Hunyuan-large: An open-source moe model with 52 billion activated parameters by tencent},
  author={Sun, Xingwu and Chen, Yanfeng and Huang, Yiqing and Xie, Ruobing and Zhu, Jiaqi and Zhang, Kai and Li, Shuaipeng and Yang, Zhen and Han, Jonny and Shu, Xiaobo and others},
  journal={arXiv preprint arXiv:2411.02265},
  year={2024}
}

@article{reynolds2011forum,
  author  = {Reynolds, Rebecca},
  title   = {Reinventing the forum: multiple perspectives, information transmission and new technology},
  journal = {Museum Management and Curatorship},
  volume  = {26},
  number  = {1},
  pages   = {45--62},
  year    = {2011},
  doi     = {10.1080/09647775.2011.540126},
  url     = {https://doi.org/10.1080/09647775.2011.540126}
}

@inproceedings{rzayev2019presence,
  author    = {Rzayev, Rufat and Karaman, G{\"u}rkan and Wolf, Katrin and Henze, Niels and Schwind, Valentin},
  title     = {The Effect of Presence and Appearance of Guides in Virtual Reality Exhibitions},
  booktitle = {Proceedings of Mensch und Computer 2019},
  series    = {MuC '19},
  pages     = {11--20},
  publisher = {ACM},
  year      = {2019},
  doi       = {10.1145/3340764.3340802},
  url       = {https://doi.org/10.1145/3340764.3340802}
}

@inproceedings{yanagawa2025navigationpixie,
  author    = {Yanagawa, Hikari and Hiroi, Yuichi and Tokida, Satomi and Hatada, Yuji and Hiraki, Takefumi},
  title     = {Navigation Pixie: Implementation and Empirical Study Toward on-Demand Navigation Agents in Commercial Metaverse},
  booktitle = {2025 IEEE International Symposium on Mixed and Augmented Reality (ISMAR)},
  pages     = {1137--1147},
  publisher = {IEEE},
  year      = {2025},
  doi       = {10.1109/ISMAR67309.2025.00120},
  url       = {https://doi.org/10.1109/ISMAR67309.2025.00120}
}

@article{feiner1997touring,
  author    = {Feiner, Steven and MacIntyre, Blair and H{\"o}llerer, Tobias and Webster, Anthony},
  title     = {A Touring Machine: Prototyping {3D} Mobile Augmented Reality Systems for Exploring the Urban Environment},
  journal   = {Personal Technologies},
  volume    = {1},
  number    = {4},
  pages     = {208--217},
  publisher = {Springer},
  year      = {1997},
  doi       = {10.1007/BF01682023},
  url       = {https://doi.org/10.1007/BF01682023}
}

@inproceedings{lopezgarcia2024speakingobjects,
  author    = {L{\'o}pez Garc{\'i}a, Irene and Schott, Ephraim and Gohsen, Marcel and Bernhard, Volker and Stein, Benno and Froehlich, Bernd},
  title     = {Speaking with Objects: Conversational Agents' Embodiment in Virtual Museums},
  booktitle = {2024 IEEE International Symposium on Mixed and Augmented Reality (ISMAR)},
  pages     = {279--288},
  publisher = {IEEE},
  year      = {2024},
  doi       = {10.1109/ISMAR62088.2024.00042},
  url       = {https://doi.org/10.1109/ISMAR62088.2024.00042}
}

@inproceedings{caetano2023arlang,
  author    = {Caetano, Arthur and Lawson, Alyssa and Liu, Yimeng and Sra, Misha},
  title     = {{ARLang}: An Outdoor Augmented Reality Application for Portuguese Vocabulary Learning},
  booktitle = {Proceedings of the 2023 ACM Designing Interactive Systems Conference},
  series    = {DIS '23},
  pages     = {1224--1235},
  publisher = {ACM},
  year      = {2023},
  doi       = {10.1145/3563657.3596090},
  url       = {https://doi.org/10.1145/3563657.3596090}
}

@inproceedings{shin2023howspace,
  author    = {Shin, Jae-Eun and Woo, Woontack},
  title     = {How Space is Told: Linking Trajectory, Narrative, and Intent in Augmented Reality Storytelling for Cultural Heritage Sites},
  booktitle = {Proceedings of the 2023 CHI Conference on Human Factors in Computing Systems},
  series    = {CHI '23},
  pages     = {241:1--241:14},
  publisher = {ACM},
  year      = {2023},
  doi       = {10.1145/3544548.3581414},
  url       = {https://doi.org/10.1145/3544548.3581414}
}

@inproceedings{trichopoulos2023llmheritage,
  author    = {Trichopoulos, Georgios},
  title     = {Large Language Models for Cultural Heritage},
  booktitle = {Proceedings of the 2nd International Conference of the ACM Greek SIGCHI Chapter (CHIGREECE 2023)},
  year      = {2023},
  publisher = {ACM},
  doi       = {10.1145/3609987.3610018},
  url       = {https://doi.org/10.1145/3609987.3610018}
}

@inproceedings{ho2025visualagents,
  author    = {Ho, Hoang Phuoc and Ramesh, Vani and Zaloudek, Ivo and Rikhtehgar, Delaram Javdani and Wang, Shenghui},
  title     = {Enhancing Visitor Engagement in Interactive Art Exhibitions with Visual-Enhanced Conversational Agents},
  booktitle = {Proceedings of the 30th International Conference on Intelligent User Interfaces (IUI '25)},
  year      = {2025},
  publisher = {ACM},
  pages     = {660--671},
  doi       = {10.1145/3708359.3712145},
  url       = {https://doi.org/10.1145/3708359.3712145}
}

@article{raptis2021mumia,
  author  = {Raptis, George E. and Kavvetsos, Giannis and Katsini, Christina},
  title   = {MuMIA: Multimodal Interactions to Better Understand Art Contexts},
  journal = {Applied Sciences},
  volume  = {11},
  number  = {6},
  pages   = {2695},
  year    = {2021},
  doi     = {10.3390/app11062695},
  url     = {https://doi.org/10.3390/app11062695}
}

@misc{pico4ultraOfficial,
  author = {{PICO}},
  title  = {{PICO 4 Ultra}},
  year   = {2026},
  url    = {https://www.picoxr.com/global/products/pico4-ultra},
  note   = {Accessed: 2026-06-24}
}

@misc{insta360Official,
  author = {{Insta360}},
  title  = {{Insta360 360 Cameras}},
  year   = {2026},
  url    = {https://www.insta360.com/consumer/360-cameras},
  note   = {Accessed: 2026-06-24}
}

@inproceedings{pan2023flow,
  author    = {Pan, Dongyijie and Li, Zhengnan and Ling, Long and Fei, Guangzheng},
  title     = {{FLOW}: Filtering and {LSTM}-based Optimization for Web Browser Interactions},
  booktitle = {Proceedings of the Eleventh International Symposium of Chinese CHI},
  series    = {CHCHI 2023},
  year      = {2023},
  pages     = {37--43},
  publisher = {Association for Computing Machinery},
  address   = {New York, NY, USA},
  doi       = {10.1145/3629606.3629611},
  url       = {https://doi.org/10.1145/3629606.3629611}
}
